\documentclass[twocolumn,showpacs,preprintnumbers,amsmath,amssymb]{revtex4}

\usepackage{graphicx}

\usepackage{dcolumn}

\usepackage{bm}
\begin{document}

%\preprint{APS/123-QED}

\title{Electron and phonon correlations in systems of one-dimensional electrons coupled to phonons}
\author{Leiming Chen}

\address{Institut f\"{u}r Theoretische Physik, Universit\"{a}t zu K\"{o}ln, Z\"{u}lpicher Strasse 77,D-50937 K\"{o}ln, Germany}

\begin{abstract}
Electron and phonon correlations in systems of one-dimensional
electrons coupled to phonons are studied at low temperatures by
emphasizing on the effect of electron-phonon backward scattering. It
is found that the $2k_F$-wave components of the electron density and
phonon displacement field share the same correlations. Both
correlations are quasi-long-ranged for a single conducting chain
coupled to one-dimensional or three-dimensional phonons, and they
are long-ranged for repulsive electron-electron interactions for a
three-dimensional array of parallel one-dimensional conducting
chains coupled to three-dimensional phonons.
%we find long-range-order charge density wave for repulsive
%electron-electron interactions in the whole parameter space of
%phonon frequency and coupling strength. We also show that a single
%one-dimensional conducting chain coupled to three-dimensional
%phonons does not lead to long-range-order charge density wave, and
%more surprisingly, even has less increased tendency to form charge
%density wave than a single conducting chain coupled to
%one-dimensional phonons.
%the Wentzel-Bardeen singularity does
%{\it not} exist in the former.
\end{abstract} 
\maketitle

\section{Introduction}

The physics of one-dimensional (1D) interacting electron systems has
attracted lots of interests. The low-energy long-wavelength
excitations of such systems can be described by a
Tomonnaga-Luttinger liquids (TLL) model  \cite{Haldane}. Such
systems exhibit a critical like behavior (power law) of the
charge-density correlations at zero temperature. Thus, 1D electron
systems are always at the verge of an instability without being to
order \cite{Giamarchi}. On the other hand, it was noticed by Peierls
\cite{Peierls} that due to electron-phonon interactions a charge
density wave (CDW) accompanied by a periodic lattice distortion
appears in the ground state of a 1D metal, both periods being
$\pi/k_F$. For an arbitrary band filling the period is
incommensurate with the underlying lattice, where the number of
electrons per site is not a simple fractional number (eg., 1/2, 1/3,
etc.). Later studies \cite{Void} on incommensurate systems of 1D
electrons coupled to 1D phonons (1D1D) show that the tendency to
form CDW is increased significantly by electron-phonon coupling;
however, true long-ranged-order CDW is still missing. In reality 1D
electron systems are often embedded in three-dimensional (3D)
environment, and phonons in such systems are effectively 3D.
Therefore, it is interesting to know if the coupling between 1D
electrons and 3D phonons could lead to long-range-order CDW.

For simplicity, we limit this paper to spinless incommensurate
systems. Our work focuses on the effect of the electron-phonon
coupling at large momentums (i.e., Backward-scattering process). We
show that the $2k_F$-wave components of the electron density and
phonon displacement field share the same correlations at large
distance. This result applies to all the systems we consider in this
paper. It is {\it exact} and does not depend on the electron-phonon
coupling strength, phonon frequency and electron-electron
interactions. This result implies that phonon correlations are
quasi-long-ranged even for arbitrary weak electron-phonon coupling
in 1D1D, in which previous studies \cite{Void} showed that electron
correlations are quasi-long-ranged.

The first system we consider is a single conducting chain coupled to
3D acoustic phonons (1D3D). We find that electron-phonon coupling in
this system does {\it not} change the qualitative power-law electron
correlations, although it does increase the tendency to form CDW
significantly (i.e., make electron correlations decay much slower).
Surprisingly, our calculation shows that 1D3D even has less
increased tendency to form CDW than 1D1D, given that both the phonon
frequency and the electron-phonon coupling strength are the same in
these two cases. Furthermore, phonon correlations are also
quasi-long-ranged in 1D3D.

The second system we discuss is a 3D array of parallel 1D conducting
chains couple to 3D acoustic phonons (3D3D).
%We treat this
%problem using two different theoretical approaches
%%developed by Voit and Schulz \cite{Void}
%based on the limits of weak coupling and/or high phonon frequency,
%and strong coupling and/or low phonon frequency. We find that,
For arbitrary weak electron-phonon coupling, the system undergoes a
quantum phase transition driven by varying electron-electron
interactions: for repulsive electron-electron interactions,
electrons on different chains are effectively correlated, which
leads to long-ranged-ordered CDW and hence long-ranged phonon
correlations; for attractive electron-electron interactions,
electron correlations remain 1D, and the system is effectively a
simple sum of 1D1D, which leads to quasi-long-ranged electron and
phonon correlations. In the limit of strong coupling and/or low
phonon frequency, both electron and phonon correlations are
long-ranged for repulsive electron-electron interactions, and
interestingly, the phase fluctuations of the $2k_F$-wave components
of the electron density and phonon displacement field are {\it not}
independent but locked together.
%We
%estimated their expectation values.
%and argue that their phase
%fluctuations are {\it always} locked together in the ordered phase
%even for arbitrary weak coupling.
%Furthermore, we notice that one of
%the requirements for a consistent description of the strong-coupling
%approach seems overlooked in the previous work on 1D1D \cite{Void}.
The 3D3D system has also been considered recently by Artemenko et al
\cite{Artemenko}. Their work agrees with ours in the limit of strong
coupling and/or low phonon frequency.

It was argued \cite{Large} that the electron-phonon coupling at
small momentums (forward-scattering process) is not important since
its effect is of the order $c^2/v_F^2\ll 1$ ($v_F$ is the Fermi
velocity, and $c$ is the sound velocity). However, it was pointed
out \cite{Thierry} that in strongly correlated systems coupling to
small-momentum acoustic phonons can lead to the Wentzel-Bardeen (WB)
singularity \cite{Wentzel,Bardeen} and hence become important. This
singularity is a critical point at which for a critical
electron-phonon coupling constant, the attractive electron-phonon
interactions cause the system to be unstable. It was shown
\cite{Thierry} that the critical point can be reached as one
approaches the half-filling for the Hubbard model
\cite{VoitHalfFilling}. In this paper we assume that the system is
far away from the WB singularity for a general filling, and the
electron-phonon coupling at small momentums is thus not important.
In fact we even find that the WB singularity does {\it not} exist in
1D3D no matter the filling. This is because the bulk phonon freedom
suppresses phonon fluctuations on the chain, thus making the system
robust.

This paper is arranged as follows. In Sec. \ref{Model} we introduce
the model. In Sec \ref{1D1D}. we review 1D1D systems. In Sec.
\ref{1D3D} and Sec. \ref{3D3D}, we discuss 1D3D and 3D3D systems,
respectively. Finally some technical details are given in the
appendixes.

\section{\label{Model}Model}
In this section we discuss the model for 3D3D, from which the models
for 1D1D and 1D3D can be obtained as special cases. Let us start
with electrons. Using the TLL description for electrons
\cite{Haldane, Giamarchi}, the electron density inside the $j_{th}$
chain can be written in the form
\begin{eqnarray}
 \rho_e^j=-{1\over\pi}\partial_x\phi_j+{1\over
 \pi\alpha}\cos\left(2k_F+2\phi_j\right)\, ,
 \label{ElectronDensity}
\end{eqnarray}
where $\phi_j$ is a slowly varying phase, $\hat{x}$ denotes the
direction along the chains, $\alpha$ is the cut-off and of the order
of the lattice spacing, $k_F$ is the Fermi momentum. The first term
on the right-hand side of (\ref{ElectronDensity}) is the
long-wavelength part of the electron density, the second term is the
fast oscillating part. The total electron action is a sum of TLL
models
\begin{eqnarray}
S_{el}/\hbar={1\over 2\pi K}\sum_j\int dxd\tau
 \left[{1\over v}(\partial_{\tau}\phi_j)^2 + v
 (\partial_x \phi_j)^2\right]\, ,
%{1\over b^2}{1\over 2\pi K}\int d^3rd\tau
% \left[{1\over v_{\rho}}(\partial_{\tau}\phi)^2 + v_{\rho}
% (\partial_x \phi)^2\right]\, ,
 \label{Electron3D1}
 \end{eqnarray}
where $v$ is renormalized Fermi velocity, $K$ is a dimensionless
parameter, which is bigger than 1 for attractive electron-electron
interactions and less than 1 for repulsive electron-electron
interactions, $\tau$ is the imaginary time.

Since electrons are restricted on the chains, an effective phonon
action in terms of the on-chain freedom is needed. It can be
obtained by starting with the standard lattice version of a 3D
phonon action and eliminating the off-chain phonon freedom. This
effective action is expected to have the following form
\begin{eqnarray}
 S_{ph}^{3D}[u]/\hbar&=&{1\over 2a\hbar }\sum_j\int dx d\tau\ \
 \left[m(\partial_{\tau}u_j)^2\right.\nonumber\\
 &&\left.+ K_xa^2(\partial_xu_j)^2
 +K_{\perp}|u_j-u_{j+1}|^2\right]\, ,
 \label{Phonon3D1}
\end{eqnarray}
where $\hbar$ is Planck Constant, $a$ is the lattice spacing, $b$ is
the distance between nearest-neighbor chains, $m$ is the atom mass,
$\perp$ denotes directions perpendicular to the chains, $K_x$ and
$K_{\perp}$ are the effective spring constants along $\hat{x}$ and
$\perp$ directions, respectively.

In general the action for electron-phonon coupling can be expressed
in the form \cite{Fetter}
\begin{eqnarray}
 S_{ep}/\hbar={\gamma\over\hbar}\sum_j\int dxd\tau\ \
 (\partial_xu_j)\rho_e^j(x)\, ,
\end{eqnarray}
%\begin{eqnarray}
%\gamma={1\over a}\int dx\ \ V(x)e^{iqx}
%\end{eqnarray}
where $\gamma$ is the coupling constant. Plugging
(\ref{ElectronDensity}) into the above action we obtain
\begin{eqnarray}
 S_{ep}/\hbar&=&{\gamma\over\pi\hbar}\sum_j\int dxd\tau\ \ \left[
 -(\partial_xu_j)(\partial_x\phi_j)+{1\over \alpha}(\partial_xu_j)\times\right.\nonumber\\
 &&\left.\cos\left(2k_Fx+2\phi_j\right)\right]
 \label{FullElectronPhonon}\, .
\end{eqnarray}
The first piece on the right-hand side of (\ref{FullElectronPhonon})
corresponds to the electron-phonon coupling at small momentums,
which is responsible for the WB singularity. We will discuss it in
section \ref{1D3D}. The second piece is the coupling at large
momentums, which is the primary focus of this paper. For this latter
type of coupling, only the phonon field with wavelength near
$1/2k_F$ is important due to the fast oscillation of the cosine
term. We write
\begin{eqnarray}
 u_j(x)={1\over 2}[e^{2ik_Fx}\tilde{\psi}_j(x)+H.c.]\, ,
 \label{Decomposition}
\end{eqnarray}
where $\tilde{\psi}_i$ is a slow varying complex field. Then the
action for the coupling at large momentums can be rewritten as
\begin{eqnarray}
S_{ep}^l/\hbar&=&{\gamma\over 4\pi\hbar}\sum_j\int dx d\tau\ \
\left(2ik_F\tilde{\psi}_j+\partial_x\tilde{\psi}_j\right)\times\nonumber\\
&&\left(e^{4ik_Fx+2\phi_j}+e^{-2i\phi_j}\right)+\mbox{H.c.}\, ,
\end{eqnarray}
where the piece which involves $e^{4ik_Fx}$ vanishes when integrated
over $x$, also $\partial_x\psi_j$ is negligible compared to
$2k_F\psi_j$ at long wavelength (i.e., $q_x<2k_F$). The action can
be further simplified as
\begin{eqnarray}
S_{ep}^l/\hbar={i\gamma k_F\over 2\pi
 \hbar\alpha}\sum_j\int dx d\tau\ \ \tilde{\psi}_j
 e^{-2i\phi_j}+\mbox{H.c.}\, .\label{ElectronPhonon3D1}
\end{eqnarray}
At this point, it is also convenient to have a phonon action in
terms of $\tilde{\psi}_j$. Plugging (\ref{Decomposition}) into
(\ref{Phonon3D1}) we obtain
\begin{eqnarray}
 S_{ph}^{3D}[\tilde{\psi}_j]/\hbar&=&{1\over 2a\hbar}\sum_j\int dx
 d\tau
 \left({m\over 2}|\partial_{\tau}\tilde{\psi}_j|^2 + 2K_xa^2k_F^2\times\right.\nonumber\\
 &&\left.|\tilde{\psi}_j|^2+{K_x\over 2}a^2|\partial_x \tilde{\psi}_j|^2
 +\right.\nonumber\\
 &&\left.{K_{\perp}\over
 2}|\tilde{\psi}_j-\tilde{\psi}_{j+1}|^2\right)
 \label{Phonon4D1}\, .
\end{eqnarray}

For convenience, we rescale the lengths and field such that the
actions are fully expressed in terms of dimensionless quantities. An
appropriate rescaling is the following
\begin{eqnarray}
 \tau&\to&a\tau /(\pi v)\, ,~~
 x\to ax/\pi\, ,\label{Rescaling1D3D1}\\
 r_{\perp}&\to& br_{\perp}/\pi\, ,~~~\psi_j=\sqrt{ m v\over 2a\hbar}\tilde{\psi}_j\, .\label{Rescaling1D3D2}
\end{eqnarray}
After the rescaling, the actions (\ref{Electron3D1}),
(\ref{Phonon4D1}) and (\ref{ElectronPhonon3D1}) become
\begin{eqnarray}
 S_{el}/\hbar&=&{1\over 2\pi K}\sum_j\int dx d\tau
 \left[(\partial_{\tau}\phi_j)^2 +(\partial_x \phi_j)^2\right]\,
 ,\nonumber\\
 \label{Electron3D4}\\
 S_{ph}^{3D}[\psi_j]/\hbar&=&{1\over 2}\sum_j\int dxd\tau\ \
 \left[|\partial_{\tau}\psi_j|^2 +g_1^2\left(1\over 2n_e\right)^2\times\right.\nonumber\\
 &&\left.|\partial_x \psi_j|^2
 +g_1^2|\psi_j|^2+\right.\nonumber\\
 &&\left.\left(C_{\perp}\over \pi^2\right)|\psi_j-\psi_{j+1}|^2
 \right]
 \label{Phonon4D3}\, ,\\
 S_{eh}^l/\hbar&=&i{g_2}\sum_j\int dxd\tau\ \ \psi_j e^{-2i\phi_j}+
 \mbox{H.c.}\, ,
 \label{ElectronPhonon3D4}
\end{eqnarray}
%\begin{eqnarray}
% S_{el}&=&{1\over 2\pi^3 K}\int d^3r dy
% \left[(\partial_y\phi)^2 +(\partial_x \phi)^2\right]\, ,
% \label{Electron3D3}\\
% S_{ph}^{3D}[\psi]&=&{1\over 2\pi^2}\int d^3rdy\ \
% \left[|\partial_y\psi|^2 +g_1^2\left(\pi\over 2ak_F\right)^2|\partial_x \psi|^2\right.\nonumber\\
% &&\left.+g_1^2|\psi|^2
% +{C_{\perp}}|\vec{\nabla}_{\perp}\psi|^2
% \right]
% \label{Phonon4D2}\, ,\\
% S_{int}&=&{g_2\over\pi^2}\int d^3rdy\left[\psi e^{-2i\phi}+
% \psi^*e^{2i\phi}
% \right]\,
% \label{ElectronPhonon3D3}
%\end{eqnarray}
where $C_{\perp} \equiv K_{\perp}a^2/m v^2$, the filling factor
$n_e=ak_F/\pi$ (i.e., the averaged number of electrons per lattice
site)\, which is of order 1 for a general filling. Both $g_1$ and
$g_2$ are dimensionless quantities defined as
\begin{eqnarray}
g_1&=& 2n_e\left(c\over v\right)=(\hbar\omega_{2k_F})\left(\hbar \pi v\over a\right)^{-1}\, ,\\
g_2&=&{\gamma\over \sqrt{2\pi}}\left(\hbar^2\over
ma^2\right)^{1\over 2}\left(\hbar\pi v\over a \right)^{-{3\over
2}}\left(an_e\over\alpha\right)\, ,
\end{eqnarray}
where $c\equiv a\sqrt{K_x/m}$ is the sound velocity along the
chains, $\omega_{2k_F}$ is the phonon frequency at wave vector
$2k_F$, $\hbar^2/ma^2$ is approximately the ground state energy of
an atom with mass $m$ trapped in an infinite one-dimensional
potential well with width $a$, $\hbar\pi v/a$ is of the order of the
Fermi energy. Typically $g_1$ is of order $10^{-5}-10^{-6}$. In the
present paper we assume
\begin{eqnarray}
 {g_2\over g_1}\ll 1\label{BasicAssumption}\, ,
\end{eqnarray}
which can be fulfilled for a general filling. This condition
actually implies that the system is far away from the WB
instability, so that the electron-phonon coupling at small momentums
is not important. For the Hubbard model very close to half-filling,
since $g_1/g_2$ which is proportional to $v^{1/2}$ drops rapidly to
zero \cite{VoitHalfFilling}, the postulates (\ref{BasicAssumption})
breaks down; therefore, the electron-phonon coupling at small
momentums could be important. This will be further discussed in
section \ref{1D3D}.

\section{\label{1D1D}A Single Conducting 1D Chain Coupled to 1D Phonons}

The model for 1D1D can be recovered by restricting the index $j$ to
be 1 in the actions (\ref{Electron3D4}), (\ref{Phonon4D3}),
(\ref{ElectronPhonon3D4}) and setting $C_{\perp}$ to be 0. To
lighten the notation we drop the index $j$ in this section. First we
want to show a very general result which is {\it exact} and does not
depend on the values of $K$, the electron-phonon coupling strength
and phonon frequency. We introduce a new complex field $\psi'$,
which, in Fourier space, is related to $\phi$ and $\psi$ by
\begin{eqnarray}
 \psi'_R(\vec{q})&=&\psi_R(\vec{q})-{2g_2G_{ph}(\vec{q})\over \sqrt{V}}\int dxd\tau\ \ \sin{2\phi}\ \ e^{i\vec{q}\cdot\vec{r}}\, ,\nonumber\\
 \label{Psi'R}\\
 \psi'_I(\vec{q})&=&\psi_I(\vec{q})-{2g_2G_{ph}(\vec{q})\over \sqrt{V}}\int dxd\tau\ \ \cos{2\phi}\ \ e^{i\vec{q}\cdot\vec{r}}\, ,\nonumber\\
 \label{Psi'I}
\end{eqnarray}
where $V\equiv L\hbar v/(k_BTa^2)$, $G_{ph}^{-1}(\vec{q})\equiv
q_{\tau}^2+g_1^2(1/2n_e)^2q_x^2+g_1^2$, the subscripts $R$ and $I$
denote the real and imaginary components, respectively. In terms of
$\psi'$ and $\phi$ the total action can be nicely separated into two
parts which are completely decoupled from each other:
\begin{eqnarray}
 S_{\psi'}/\hbar&=&\sum_{\vec{q}}G_{ph}^{-1}(\vec{q})\psi'(\vec{q})\psi'(-\vec{q})\, ,\\
 S_{el}^{1D}/\hbar&=&{1\over 2\pi K}\int dx d\tau\ \
 \left[(\partial_{\tau}\phi)^2 + (\partial_x \phi)^2\right]
 \nonumber\\
 &~&-{g_2^2\over g_1}
 \int dx d\tau d\tau'\ \ G_{ph}(\tau-\tau')
 \cos\left[2\phi(x, \tau)-\right.\nonumber\\
 &&\left.2\phi(x, \tau')\right]\, ,
 \label{EffectElectron1D1}
\end{eqnarray}
where the phonon correlation function is given by
\begin{eqnarray}
 G_{ph}(\tau)
 \approx e^{-g_1|\tau|}\, .
\end{eqnarray}
To obtain (\ref{EffectElectron1D1}) we have neglected the phonon
dispersions $(\partial_x\phi)^2$ since the leading order effect on
electron correlations comes from the phonon kinetic energy
$(\partial_{\tau}\phi)^2$. Since $\psi'$ and $\phi$ are decoupled,
from (\ref{Psi'R}) and (\ref{Psi'I}) the fluctuations of $\psi$ can
be calculated as
\begin{eqnarray}
 \langle\psi(\vec{q})\psi(-\vec{q})\rangle&=&{16\pi^2\alpha^2g_2^2G_{ph}^2
 (\vec{q})}\int dxd\tau\ \ \left[e^{i\vec{q}\cdot\vec{r}}\times\right.\nonumber\\
&&\left.\langle\rho_{2k_F}(\vec{r})\rho_{2k_F}(\vec{0})\rangle\right]
+G_{ph}(\vec{q})
 \, ,\label{GeneralRelation}
\end{eqnarray}
where $\rho_{2k_F}$ is the $2k_F$-wave component of the electron
density. Since $G_{ph}(\vec{q})$ tends toward a constant for $q\ll
g_1$, this result implies that at large distance (i.e., $r\gg
g_1^{-1}$) the $2k_F$-wave component of the phonon displacement
field has the same correlations as $\rho_{2k_F}$.

\begin{figure}
\includegraphics[width=0.35\textwidth,angle=-90]{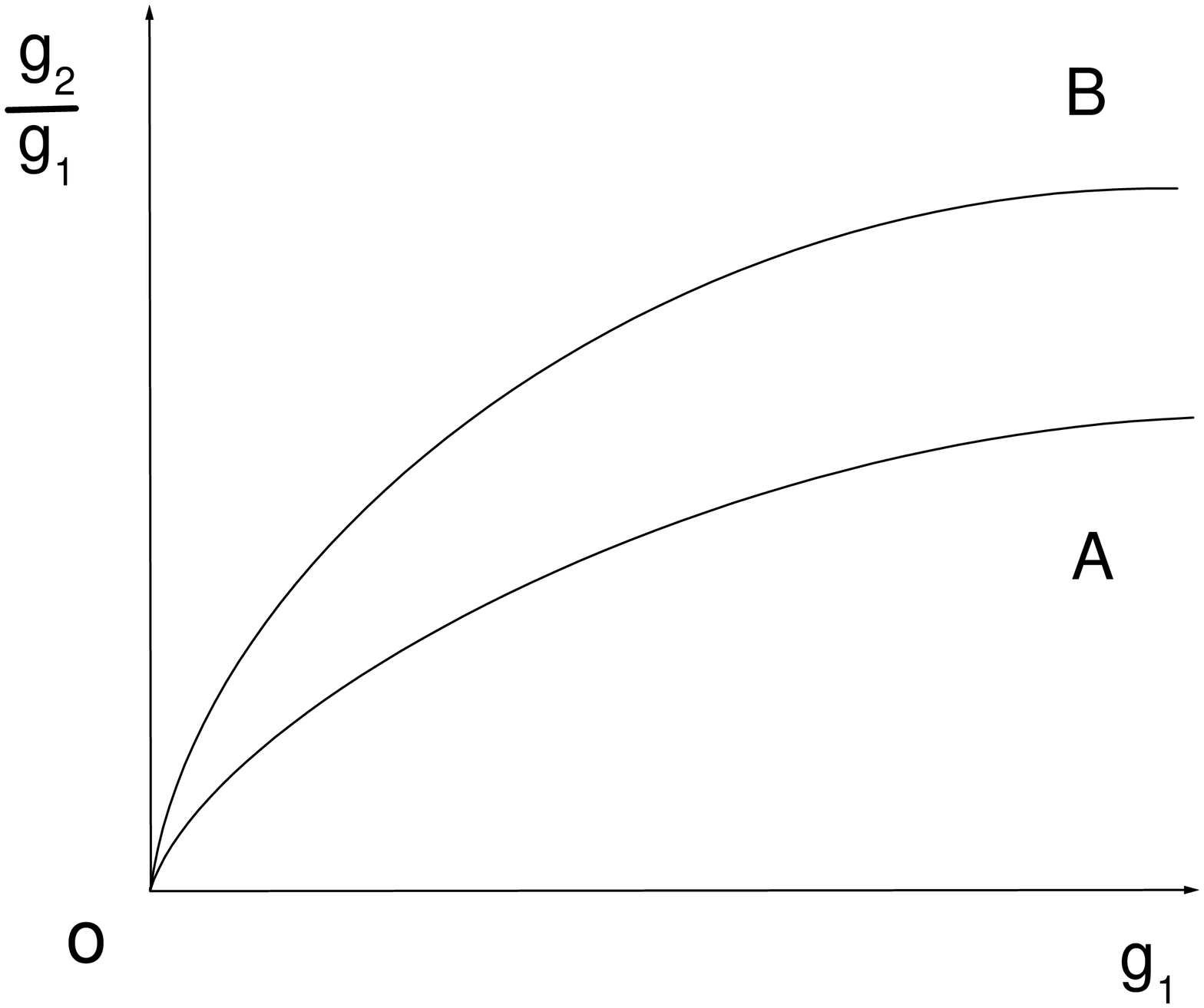}
\caption{\label{fig: StrongWeak}Limits of the validity of the two
approaches. The result of the Gaussian variational method does not
apply in the region above $OA$, while the fixed-amplitude
approximation does not apply in the region below $OB$. The regions
below $OA$ and above $OB$ are defined by the conditions
(\ref{WeakCondition}) and (\ref{StrongCondition}) in the 1D1D case and by (\ref{WeakCondition3D}) and
(\ref{StrongCondition3D}) in the 3D3D case, respectively. In the intermediate
region between $OA$ and $OB$ neither of the two approaches
applies.}
\end{figure}

To understand the detailed behavior of the correlations, more
calculations are needed. In the limit of weak coupling and/or high
phonon frequency (i.e., $g_2g_1^{K-2}\ll 1$), the effort is put on
the phonon-mediated effective electron action given by
(\ref{EffectElectron1D1}). Since the exponentially decaying phonon
correlation $G_{ph}(\tau)$ imposes a cutoff at $\tau\sim 1/g_1$, at
the wavelength longer than $1/g_1$ the model
(\ref{EffectElectron1D1}) essentially reduces to the one for free
electrons but with renormalized $K$, which we denote as $K'$. A
Gaussian variational method \cite{Orignac} and a perturbative
renormalization group (RG)\cite{Void, Orignac} have been applied to
this model. Both find that electron-phonon coupling increases the
tendency to form CDW. Specifically, in terms of $K'$ the result is
expressed as
\begin{eqnarray}
 \left(K\over K'\right)^2-1\sim \left\{\begin{array}{ll}
 \left[\left(g_2\over g_1\right)g_1^{K-1}\right]^2\, &K<{3\over 2}\\
 -{g_2^2\over g_1}\ln{g_1}\, &K={3\over 2}\\
 {g_2^2\over g_1}\, &K>{3\over 2}
 \end{array}\right.\, .
 \label{K1}
\end{eqnarray}
The details of the calculation are given in appendix \ref{HW1D1D}.
To obtain the above result, it has been assumed that the
renormalization of $K$ is small, that is, $(K/K')^2-1\ll 1$, which
leads to a self-consistent condition. For $K<1$, this
self-consistent condition is just
\begin{eqnarray}
{g_2\over g_1}\ll g_1^{1-K}\, , \label{WeakCondition}
\end{eqnarray}
which prohibits the application of the Gaussian variational result
in the region above the locus $OA$ in Fig. \ref{fig: StrongWeak}.

When the condition (\ref{WeakCondition}) is violated, neither the
Gaussian variational method nor the perturbative RG can give an
analytical result. A new strategy is used. A mean-field theory shows
that $\psi$ saturates for $K<1$. Therefore, it is believed that for
$K<1$ and in the limit of strong coupling and/or low phonon
frequency, the important fluctuations of $\psi$ are its phase
fluctuations, and the amplitude fluctuations are massive and hence
negligible. Based on this belief we now work on the truncated
action:
%This strategy
%assumes that the amplitude of $\psi$ saturates locally and is
%non-fluctuating, the important fluctuations of $\psi$ are its phase
%fluctuations, so we can obtain a simplified model by writing
%$\psi=\psi_0e^{i\theta}$ with $\psi_0$ fixed at the saturated value:
\begin{eqnarray}
 S_{total}^{1D}/\hbar&=&
 {1\over 2}\int dx d\tau \left\{{1\over\pi K}
 \left[(\partial_{\tau}\phi)^2 +
 (\partial_x \phi)^2\right]+\psi_0^2\times\right.\nonumber\\
 &&\left.\left[(\partial_{\tau}\theta)^2
 +g_1^2\left(1\over 2n_e\right)^2
 (\partial_x \theta)^2\right]\right\}\nonumber\\
 &~&-
 2g_2\psi_0\int dx d\tau
 \cos\left(2\phi-\theta\right)\, ,
 \label{Total1D1D}
\end{eqnarray}
where $\psi_0$ is fixed at the saturated value, and the constant
pieces have been thrown away. Hereafter we name this strategy as
``fixed-amplitude approximation'' just for convenience. In the
following we will first estimate the saturated value of $\psi_0$,
then calculate the fluctuations of $\phi$ and $\theta$ from the
truncated model (\ref{Total1D1D}), and finally discuss the limits of
the validity of the fixed-amplitude approximation.

We estimate the saturated value of $\psi$ using a mean-field theory.
Approximating $\psi$ as uniform both in space and time by doing
$\psi\equiv\psi_0$, the total action reduces to
\begin{eqnarray}
 S_{total}^{1D}/\hbar
 &=&{1\over 2}\int dx d\tau \ \ {1\over\pi K}
 \left[(\partial_{\tau}\phi)^2 +
 (\partial_x \phi)^2\right]\nonumber\\
 &&-2g_2\psi_0\int dx d\tau\ \ \cos\left(2\phi\right)
 \nonumber\\
 &&+{1\over 2}\int dx d\tau\ \ g_1^2\psi_0^2\, ,
 \label{MeanField1D1D}
\end{eqnarray}
where the electron part of the action is just a Sine-Gordon model.
Directly integrating out the field $\phi$ in favor of $\psi_0$ using
a Gaussian variational method, we get
\begin{eqnarray}
S_{ph}^{eff}[\psi_0]/\hbar&=&{g_1^2\over 2}\int dx d\tau
 \left[2\pi K\left(2\pi Kg_2g_1^{K-2}\psi_0\right)^
 {2\over {2-K}}\times\right.\nonumber\\
 &&\left.(K-2)+\psi_0^2\right]
 %S_{ph}^{1D}[\psi_0]&=&{1\over 2\pi K}\int dx dy
% \left[-(2-K)\left(2\pi Kg_2\phi_0\right)^
% {2\over {2-K}}
% +K\left(2\pi Kg_2\phi_0\right)^{4\over{2-K}}
% \right]\nonumber\\
% &~&+{1\over 2}\int dx dy\ \ g_1^2\psi_0^2\, .
 \, .
 \label{MeanFieldPhonon}
 \end{eqnarray}
The detailed calculation is shown in appendix \ref{1D1DStrongLow}.
The first term in the action (\ref{MeanFieldPhonon}) is negative,
which means that a periodic lattice distortion with period $\pi/k_F$
lowers the electron energy; the second term is the elastic energy
for such a lattice distortion and hence positive. To find the ground
state, we minimize the action (\ref{MeanFieldPhonon}) with respect
to $\psi_0$. We find
\begin{eqnarray}
 \psi_0=g_1^{-1}\left[2(\pi K)^{K\over 2}\left(g_2\over g_1\right)
 \right]^{1\over 1-K}
 \label{Solution1}\, \label{Psi}
\end{eqnarray}
for $K<1$ and $\psi_0=0$ for $K>1$. This indicates that a periodic
lattice distortion with period $\pi/k_F$ appears in the ground state
for repulsive electron-electron interactions; however, $\psi$ is
{\it not} long-range-ordered due to its phase fluctuations. Later in
this section we will show that $\psi$ is quasi long-range-ordered
with power-law correlations.
%In the limit of
%no quantum fluctuations (i.e., $K\to 0$), (\ref{Psi}) leads to
%$\psi_0=2g_2g_1^{-2}$.
For $K$ not very close to 1, the result given in (\ref{Psi})
qualitatively agrees with that in ref. \cite{Void}, in which the
electron freedom is eliminated by mapping the model
(\ref{MeanField1D1D}) to an exactly solvable field-theoretical model
(i.e., Massive Thirring model) and calculating the ground state
energy of electrons. In the limit of $K\to 1^-$, (\ref{Psi}) shows
$\psi_0\to 0$, and the classic Peierls theory predicts $\psi_0\sim
\exp\left(-g_1^2/g_2^2\right)/g_2$, which is essentially very small.

The above mean-field calculation assumes that the single electron
gap opened at the Fermi level is much less than the Fermi energy,
that is, $(g_2\psi_0)^{1/(2-K)}\ll 1$. This assumption combined with
(\ref{Psi}) leads to the postulated condition
(\ref{BasicAssumption}), or equivalently
\begin{eqnarray}
 g_1\psi_0\ll 1\label{ImplicitCondition}\, .
\end{eqnarray}
\begin{figure}
\includegraphics[width=0.35\textwidth,angle=-90]{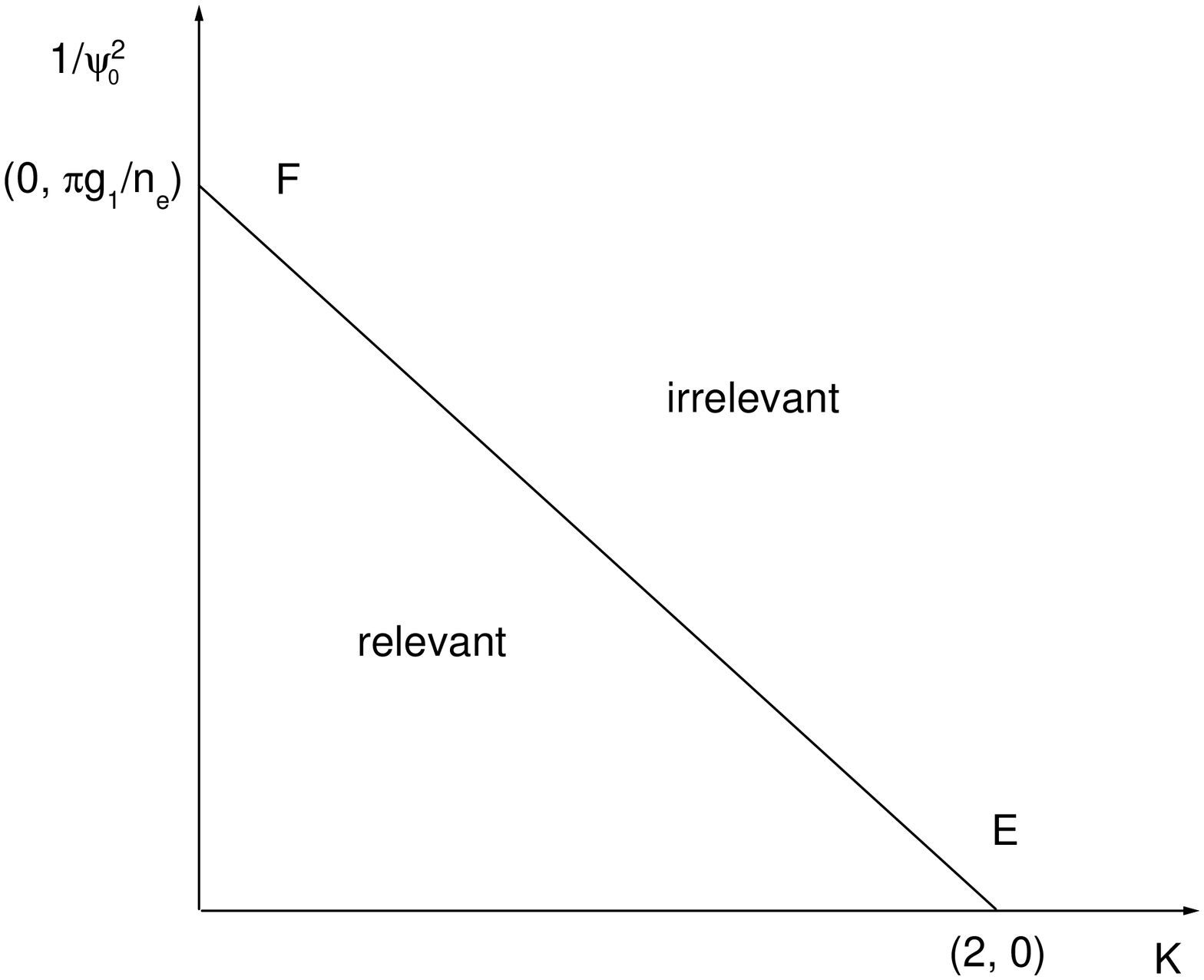}
\caption{\label{fig: CriticalLine}The line $EF$ is defined by the
function given in (\ref{CriticalLine}). The cosine term in the
action (\ref{Total1D1D}) is irrelevant above $EF$ and relevant below
it.}
\end{figure}
%For the approximation that $\psi_0$ is non-fluctuating to work,
%certain conditions must be satisfied. One condition comes directly
%from the restrictions on the amplitude fluctuations
%$\langle(\delta\psi_0)^2\rangle\ll\psi_0^2$. We go over the
%calculation in appendix \ref{1D1DStrongLow}.
%Here we will show that another condition due to the phase
%fluctuations of $\psi$ can also lead to inconsistency of the
%strategy. Specifically, Now let's calculate $\psi_0^c$.

Now we discuss the truncated model (\ref{Total1D1D}). This model has
been studied previously in the context of liquid crystal
\cite{Nelson}. It can be shown that the coupling term in this model
is either relevant or irrelevant depending on the values of $\psi_0$
and $K$. The critical values of $\psi_0$ and $K$ can be easily
calculated in the limit $K\to 0$ (i.e., the electron field $\phi$ is
locked) or $\psi_0\to\infty$ (i.e., the phonon phase field $\theta$
is locked) since in both limits the model reduces to the Sine-Gordon
model. We get $K_c=2$ for $\psi_0\to\infty$ and
$\psi_0^c=\sqrt{n_e/\pi g_1}$ for $K\to 0$. In general, the critical
values of $\psi_0$ and $K$ satisfies
\begin{eqnarray}
 {n_e\over\pi g_1 \psi_0^2}+2K=4\, .
 \label{CriticalLine}
\end{eqnarray}
The detailed derivation is given in appendix \ref{PsiC}. This
function defines a critical line in the $\psi_0^{-2}-K$ plane, which
is illustrated in Fig. \ref{fig: CriticalLine}. Thus, for $K<1$,
only for which we can get a non-zero $\psi_0$, we have
\begin{eqnarray}
\psi_0^c\sim g_1^{-{1\over 2}}\, . \label{PsiCritical}
\end{eqnarray}
For $\psi_0<\psi_0^c$, the cosine term becomes irrelevant, and
$\phi$ and $\theta$ are decoupled. For $\psi_0>\psi_0^c$, the cosine
term is relevant, and the fluctuations of $\phi$ and $\theta$ are
bound together. This actually implies a self-consistent condition
for the fixed-amplitude approximation; that is, the estimated
$\psi_0$ must be bigger than $\psi_0^c$ such that phase fluctuations
do not drive the cosine term into irrelevance. This cosine term
offsets the positive phonon energy and is necessary for obtaining a
non-zero $\psi_0$ in the first place. In terms of $g_1$ and $g_2$
this self-consistent condition
%to phase fluctuations is
%\begin{eqnarray}
%\psi_0\gg g_1^{-{1\over 2}}\label{Condition3}
%\end{eqnarray}
%which
is given by
\begin{eqnarray}
{g_2\over g_1}\gg g_1^{1-K\over 2}\, , \label{StrongCondition}
\end{eqnarray}
which prohibits the using of the fixed-amplitude approximation in
the region below the locus $OB$ in Fig. \ref{fig: StrongWeak}.
%This condition seems identical to the one obtained from the
%amplitude fluctuations; however, there is a crucial difference
%between these two. When we derive (\ref{Condition3}) from the phase
%fluctuations, the $g_1$ on the right side comes from the coefficient
%of $(\partial_x\psi)^2$ which is scaling invariant under the RG up
%to one loop. The RG flow equations are given in appendix
%\ref{HW1D1D}. Using the expression (\ref{Solution1}), $\psi_0$ is
%also found to be scaling invariant. As a result the ``phase version"
%of (\ref{StrongCondition}) is scaling invariant overall and
%prohibits the strong-coupling strategy from being applied to the
%systems whose bare parameters do not satisfy
%(\ref{StrongCondition}). On the other hand, when we derive
%(\ref{Condition3}) from the amplitude fluctuations, the $g_1$ on the
%right side comes from the coefficient of $\psi^2$ and thus scales as
%$g_1e^{\ell}$. Under the RG the ``amplitude version" of
%(\ref{StrongCondition}) scales as
%\begin{eqnarray}
% \left(g_2^2g_1^{K-3}\right)^{1\over 2(1-K)}e^{\ell\over 2}\gg 1
%\end{eqnarray}
%which becomes easily fulfilled for large $\ell$. Therefore the
%self-consistent condition due to the phase fluctuations is stronger.
For the systems satisfying (\ref{StrongCondition}) both the
effective actions of $\phi$ and $\theta$ can be obtained by making
$2\phi=\theta$ in the truncated model (\ref{Total1D1D}), and they are
given by
\begin{eqnarray}
 S[\Xi]/\hbar&=&{1\over 2\pi K}\int dxd\tau\left[R(\partial_{\tau}\Xi)^2
 +B(\partial_x\Xi)^2\right]\, ,
 \label{Fukuyama1}
\end{eqnarray}
where $\Xi$ can be either $\phi$ or $\theta/2$, and the parameters
$R$ and $B$ are
\begin{eqnarray}
 R&=&(1+4\pi K\psi_0^2)\label{Strong1DR}\, ,\\
 B&=&1+\pi Kg_1^2\left(1\over n_e\right)^2\psi_0^2
 \label{Strong1DB}\, .
\end{eqnarray}
This result is consistent with our earlier conclusion that at large
distance the $2k_F$-wave components of the phonon displacement field
and electron density share the same correlations. Note that the
correction to $B$ comes from the phonon dispersions along the chains
and is much less than 1 according to the assumption
(\ref{ImplicitCondition}). Also it is of the order of the correction
to $R$ multiplied by $g_1^2$ and hence negligible. Plugging
(\ref{Psi}) into (\ref{Strong1DR}) and defining $K'\equiv
K/\sqrt{R}$, we get
\begin{eqnarray}
 \left(K\over K'\right)^2-1 &\sim& g_1^{-2}\left(g_2\over g_1\right)^{2\over
 (1-K)}\, .
\end{eqnarray}

Now let us discuss the validity of the fixed-amplitude
approximation. In addition to the self-consistent condition (\ref{StrongCondition})
caused by phase fluctuations, there is another one caused by amplitude fluctuations,
which is
\begin{eqnarray}
 \langle(\delta\psi_0)^2\rangle\ll \psi_0^2\, ,
 \label{AmplitudeCondition}
\end{eqnarray}
where $\delta\psi_0$ is the deviation of the amplitude of $\psi$
from the saturated value. For a crude estimation of
$\langle(\delta\psi_0)^2\rangle$, we assume that the effective
action for $\delta\psi_0$ (to the quadratic order) has the following
form
\begin{eqnarray}
 S[\delta \psi_0]/\hbar&=&{1\over 2\pi^2}\int dx d\tau\ \
 \left[{{2(1-K)}g_1^2\over {2-K}}(\delta \psi_0)^2 +\right.\nonumber \\
 &&\left.(\partial_{\tau} \delta\psi_0)^2
 +g_1^2\left(1\over
 2n_e\right)^2(\partial_x\delta\psi_0)^2
 \right]
 \, ,
 \label{ExpansionAroundSaddlePoint1D}
\end{eqnarray}
where the mass term is obtained by expanding the mean-field action
(\ref{MeanFieldPhonon}) around its minimum, and the derivative terms
come from the free phonon action. Then after a straightforward
calculation, the condition (\ref{AmplitudeCondition}) also leads to
(\ref{StrongCondition}) coincidentally.
%\begin{eqnarray}
% \left\{\begin{array}{ll}
% {g_2\over g_1}\gg g_1^{1-K\over 2}\, ,&C_{\perp}\ll g_1^2\, ,\\
% {g_2\over g_1}\gg g_1^{1-K\over 2}\left(g_1\over\sqrt{C_{\perp}}\right)^{1-K\over
% 2}\, ,&C_{\perp}\gg g_1^2\, ,
% \end{array}
% \right.
%\end{eqnarray}
%which prohibits the application of the fixed-amplitude approximation
%in the region below $OB$($OD$) for $C_{\perp}\ll
%g_1^2$($C_{\perp}\gg g_1^2$) in figure \ref{fig: StrongWeak3D}.

In both limits the charge-density correlations can be calculated,
given by the non-universal power laws
\begin{eqnarray}
 \langle\rho_{2k_F}(\vec{r})\rho_{2k_F}(0)\rangle\sim r^{-2K'}\, ,
\end{eqnarray}
where $\vec{r}=(x, \tau)$. This implies that in electron-phonon
coupling does {\it not} change the qualitative TLL behavior,
although the tendency to form CDW is increased since $K'$ is smaller
than $K$. Based on the general conclusion given in
(\ref{GeneralRelation}), at large distance phonon correlations are
also governed by the same non-universal power laws:
\begin{eqnarray}
 \langle\psi(\vec{r})\psi(0)\rangle\sim r^{-2K'}\, .
\end{eqnarray}
Surprisingly, this implies that in 1D1D while electron-phonon
coupling does not have a qualitative effect on electron
correlations, it does change phonon correlations qualitatively, that
is, the correlations of the $2k_F$-wave component of the phonon
displacement field, which are short-ranged in the absence of
electron-phonon coupling, become quasi long-ranged. According to the
self-consistent conditions given by (\ref{WeakCondition}) and
(\ref{StrongCondition}), there is an intermediate region in which
neither of the two approaches apply, and therefore no quantitative
result is available. This region is illustrated in figure \ref{fig:
StrongWeak} as the area between the loci $OA$ and $OB$. However, in
this region we do expect the same qualitative conclusion.

\section{\label{1D3D}A Single Conducting 1D Chain Coupled to 3D Phonons}

In this section we consider a single 1D conducting chain coupled to
3D phonons. In this case the action for electrons and
electron-phonon coupling is the same as that in 1D1D, and the phonon
action is given in (\ref{Phonon3D1}).
For convenience, we transform the phonon action into its continuum
version and perform the dimensional rescaling given in
(\ref{Rescaling1D3D1}, \ref{Rescaling1D3D2}) and the field rescaling
\begin{eqnarray}
 u'=\sqrt{ m v\over 2a\hbar}u\, .
 \label{Rescaling1D3D3}
\end{eqnarray}
After this procedure, we obtain
\begin{eqnarray}
 S_{ph}^{3D}[u']/\hbar&=&{1\over \pi^2}\int d^3r d\tau\ \
 \left[(\partial_{\tau}u')^2 + g_1^2\left(1\over 2n_e\right)^2\times\right.\nonumber\\
 &&\left.(\partial_xu')^2
 +C_{\perp}|\vec{\nabla}_{\perp}u'|^2\right]\, .
 \label{}
\end{eqnarray}
Since electron-phonon coupling only happens on the chain, an
effective phonon action for the on-chain freedom is needed. We
integrate out the bulk off-chain phonon freedom in favor of the
on-chain one $u'_0(x, \tau) \equiv u'(x,\tau,
\vec{r}_{\perp}=\vec{0})$, where we have assumed that the position
of the chain is at $r_{\perp}=0$. This leads to an effective 1D
phonon action
\begin{eqnarray}
 S_{ph}^{1D}[u_0']/\hbar={1\over 2}\sum_{\vec{q}}G^{-1}(\vec{q})
 |u'_0(\vec{q})|^2\, ,
 \label{Contribution1}
\end{eqnarray}
where
\begin{eqnarray}
 G(\vec{q})={\pi\over 8C_{\perp}}\ln{{q_{\tau}^2+{4\over\pi}C_{\perp}
 +\left(g_1\over 2n_e\right)^2q_x^2}\over {q_{\tau}^2+\left(g_1\over 2n_e\right)^2q_x^2}}\, .
 \label{NewPhononPropogator}
\end{eqnarray}
In the limit $C_{\perp}\to 0$, the renormalized 1D phonon propagator
given by (\ref{NewPhononPropogator}) reduces to the bare 1D one
(i.e., the one which is not renormalized by the bulk phonon freedom)
as expected. For $C_{\perp}$ being non-zero, for small $\vec{q}$s
the renormalized one has a logarithmic dependence on $\vec{q}$,
while the bare one has a second-order power-law dependence. This
difference has a drastic effect on the electron-phonon coupling at
small momentums. We delay this discussion to the later of this
section.
%we follow reference
%\cite{Thierry} integrating out the small-momentum phonon freedom.
%This leads to a {\it negative} contribution to the electron action.
%In 1D3D this contribution is, however, always sub-dominant to the
%original electron action. As a result, the WB singularity, which
%exists in 1D1D, does {\it not} exist in 1D3D. In 1D1D, a critical
%electron-phonon coupling makes the negative contribution comparable
%to the original electron action, thus making the system unstable. In
%1D3D, the bulk phonon freedom suppresses phonon fluctuations along
%the chain, thus making the system more robust.
Now we focus our attention on the electron-phonon coupling at large
momentums (i.e., near $2k_F$). Expanding the action
(\ref{Contribution1}) around $q_x=\pm 2ak_F/\pi, q_{\tau}=0$ (the
factor $a/\pi$ is due to the rescaling), we get
\begin{eqnarray}
 S_{ph}^{1D}[\psi]/\hbar={1\over 4}\sum_{q_x q_{\tau}}G^{-1}\left({2ak_F\over\pi}+q_x, q_{\tau}\right)
 |\psi (q_x, q_{\tau})|^2\, ,\nonumber\\
 \label{Contribution2}
\end{eqnarray}
where the complex phonon field $\psi$ is the slow varying part of
the $2k_F$-wave component of the displacement field $u'_0$
\begin{eqnarray}
 u'_0(x)={1\over 2}[e^{2ik_Fx}\psi(x)+H.c.]\, .
\end{eqnarray}
The inverse of the propagator in (\ref{Contribution2}), to the
lowest order of $q_{x, \tau}$, is given by
\begin{eqnarray}
 G^{-1}\left({2ak_F\over\pi}+q_x, q_y\right)
 %&&={8C_{\perp}\over\pi}
% \left(\ln{b \over a}\right)^{-1}
% \left(1+Mq_y^2+Nq_x^2\right)
 =2\left({g_1'}^2+g_3q_y^2+g_4q_x^2\right)
 \label{Propagator Contribution2}
\end{eqnarray}
with
\begin{eqnarray}
 g'_1&\equiv& 2\sqrt{{C_{\perp}\over\pi}\left(\ln{{g_1^2+4C_{\perp}/\pi} \over g_1^2}\right)^{-1}}\, ,\\
 g_3&\equiv& {{4{g'_1}^2C_{\perp}/\pi}\over{g_1^2(g_1^2+4C_{\perp}/\pi)}}
 \left(\ln{{g_1^2+4C_{\perp}/\pi} \over g_1^2}\right)^{-1}
 \, ,\\
 g_4&\equiv& {\pi^2g_1^2g_3\over 4ak_F^2}\left[1+4g_1^2\left({g_3\over {g_1'}^2}
 -{1\over {g_1^2+4C_{\perp}/\pi}}\right)\right]\, .
\end{eqnarray}
To get (\ref{Propagator Contribution2}), we have thrown away the
term linear in $q_x$ (i.e., $q_x|\psi|^2$), which only leads to
boundary effects. Plugging (\ref{Propagator Contribution2}) into
(\ref{Contribution2}) we get
\begin{eqnarray}
 S_{ph}^{1D}[\psi]/\hbar&=&{1\over 2}\int dx d\tau\ \
 \left[g_3|\partial_{\tau}\psi|^2 +
 {g'_1}^2|\psi|^2+\right.\nonumber\\
 &&\left.g_4|\partial_x \psi|^2\right]\, .
 \label{EffectivePhonon1D3D}
\end{eqnarray}
Therefore, with respect to the electron-phonon coupling at large
momentums, the effective model of phonons in 1D3D is qualitatively
the same as that in 1D1D but with modified coefficients. Thoughtful
reader might have seen this right from the beginning \cite{Smart
ass}.
%with
%\begin{eqnarray}
% g'_1&\equiv& 2\sqrt{{C_{\perp}\over\pi}\left(\ln{b \over a}\right)^{-1}}\, ,\\
% g_3&\equiv& M{g'_1}^2\, ,\\
% g_4&\equiv& N{g'_1}^2\, .
%\end{eqnarray}
%In the limit $C_{\perp}\to 0$, the action
%(\ref{EffectivePhonon1D3D}) reduces to (\ref{Phonon4D3}), which
%provides a well check of our calculation.
Note that $g'_1$, $g_3$
and $g_4$ are all increasing functions of $C_{\perp}$. This is
expected since the bulk phonon freedom suppresses phonon
fluctuations on the chain.

Since we have shown that the models for 1D3D and 1D1D are essentially
the same, the calculations should be also very similar. Repeating
the (virtually same) calculation, we obtain, for 1D3D and repulsive electron-electron interactions,
\begin{eqnarray}
 \left(K\over K'\right)^2-1\sim {g_2^2\over g'_1}\left(\sqrt{g_3}\over g'_1\right)^{3-{2K}}
 \label{}
\end{eqnarray}
in the limit of weak coupling and/or high phonon frequency and
\begin{eqnarray}
 \left(K\over K'\right)^2-1\sim g_3\left(g_2^2
 {g'_1}^{2K-4}\right)^{1\over {1-K}}
 \label{}
\end{eqnarray}
in the limit of strong coupling and/or low phonon frequency.
%The
%mean-field prediction of the saturated value of $\psi$ for $K<1$ is given by
%\begin{eqnarray}
% \psi_0=\left[2(\pi K)^{K\over 2}g_2{g_1'}^{K-2}\right]^{1\over
% {1-K}}
% \label{}\, ,
%\end{eqnarray}
%which is smaller than the one in 1D1D since $g_1'$ is larger than
%$g_1$.
It can be verified that both $K'$ obtained from the two limits are
increasing functions of $C_{\perp}$. Therefore, we conclude that a
single 1D chain coupled to 3D phonons does not lead to
long-range-ordered CDW, instead the increased tendency to form CDW
in such systems is even less than that in 1D1D, given that the two
dimensionless quantities $g_1$ and $g_2$ (or equivalently, the
phonon frequency and the coupling strength) are the same in these
two cases. This result does not seem obvious just from qualitative
arguments. On one hand, the bulk phonon freedom increases the energy
cost to form a lattice distortion along the chain, which is
manifested in that $g'_1$ is a increasing function of $C_{\perp}$.
This is a negative effect on the increased tendency to form CDW. On
the other hand, both $g_3$ and $g_4$ are increasing functions of
$C_{\perp}$, which enhances the correlations of $\psi$. This is a
positive effect on the increased tendency to form CDW. Only a
quantitative calculation can tell that the former effect dominates.
In principle we could also think about a single 1D chain coupled to
2D phonons. We expect that in such systems long-range-ordered CDW is
still missing, and the increased tendency to form CDW is less than
that in 1D1D but more than that in 1D3D, given that both $g_1$ and
$g_2$ are the same in the three cases. In addition, the $2k_F$-wave
components of the phonon displacement field in 1D2D and 1D3D are
quasi long-range-ordered due to electron-phonon coupling.

Now we discuss the electron-phonon coupling at small momentums.
According to the action given in (\ref{FullElectronPhonon}), the
action for the electron-phonon coupling at small momentums is given
by
\begin{eqnarray}
 S_{ep}^s/\hbar=-{\gamma\over\pi\hbar}\int dxd\tau\ \
 (\partial_xu_0)(\partial_x\phi)\, .
 \label{}\,
\end{eqnarray}
After the dimensional and field rescaling which are given in
(\ref{Rescaling1D3D1}) and (\ref{Rescaling1D3D3}) respectively, the
above action becomes
\begin{eqnarray}
 S_{ep}^s/\hbar=-{2\pi\alpha g_2\over an_e}\int dxd\tau\ \
 (\partial_xu_0')(\partial_x\phi)
 \label{}\, .
\end{eqnarray}
The effective action for the on-chain phonon field is given by
(\ref{Contribution1}). Since the total action is quadratic in
$u_0'$, we integrate it out. This leaves us an effective electron
action
\begin{eqnarray}
 S_{el}^{total}&=&{1\over 2\pi
 K}\sum_{\vec{q}}\left[q_x^2+q_{\tau}^2-{4\pi^3
 K\left(\alpha g_2\over an_e\right)^2q_x^4}G(\vec{q})\right]\nonumber\\
 &&\times\phi(-\vec{q})\phi(\vec{q})\, ,
 \label{1DEffectiveSmallElectron}
\end{eqnarray}
where $G(\vec{q})$ is the phonon propagator. The first two terms on
the right-hand side of (\ref{1DEffectiveSmallElectron}) are from
free electrons, and the third term, which is retarded and {\it
negative}, is contributed by phonons. First let us treat phonons as
pure 1D. Plugging $G(\vec{q})$ as the bare 1D phonon propagator into
(\ref{1DEffectiveSmallElectron}), we get
\begin{eqnarray}
 S_{el}^{total}&=&{1\over 2\pi
 K}\sum_{\vec{q}}\left[q_x^2+q_{\tau}^2-{2\pi^3
 K\alpha^2g_2^2q_x^4\over{a^2n_e^2\left[q_{\tau}^2+\left(g_1\over 2n_e\right)^2q_x^2\right]}}\right]
 \nonumber\\
 &&\times\phi(-\vec{q})\phi(\vec{q})\,
 .
 \label{1D1DEffectiveSmallElectron}
\end{eqnarray}
In the limit $q_{\tau}\to 0$, the retarded part is proportional to
$q_x^2$ and cancels the first term at the point
\begin{eqnarray}
 {1\over K}=8\pi^3\left(\alpha g_2\over ag_1\right)^2
 %={\gamma^2\over \pi\hbar c^2(m/a)}
 \, ,
 \label{WB}
\end{eqnarray}
at which the electron density becomes unstable towards long
wavelength fluctuations. This singular point is referred to as the
Wentzel-Bardeen (WB) singularity. Since $K$ and $\alpha/a$ are both
of order 1, as long as the postulates (\ref{BasicAssumption}) is
fulfilled, the left-hand side of (\ref{WB}) is much larger than the
right-hand side, and the electron-phonon coupling at small momentums
can thus be safely neglected. As we mentioned earlier, for the
Hubbard model near half-filling, the postulates
(\ref{BasicAssumption}) breaks down, and the WB singularity is
reachable; however, in this case the umclapp effect becomes also
important, and whether the WB singularity is robust against this
additional effect remains an open question. Now we treat phonons as
3D. Plugging (\ref{NewPhononPropogator}) into
(\ref{1DEffectiveSmallElectron}), we obtain
\begin{eqnarray}
 S_{el}^{total}&=&{1\over 2\pi
 K}\sum_{\vec{q}}\left[q_x^2+q_{\tau}^2-{{\pi^4
 K\over 2C_{\perp}}\left(\alpha g_2\over an_e\right)^2q_x^4}\times\right.\nonumber\\
 &&\left.\ln{{q_{\tau}^2+{4\over\pi}C_{\perp}
 +\left(g_1\over 2n_e\right)^2q_x^2}\over {q_{\tau}^2+\left(g_1\over 2n_e\right)^2q_x^2}}\right]\phi(-\vec{q})\phi(\vec{q})\,
 .\nonumber\\
 \label{3DEffectiveSmallElectron}
\end{eqnarray}
For small $\vec{q}$s the negative retarded part in
(\ref{3DEffectiveSmallElectron}) is of the order $q_x^4$ with a
logarithmic correction, which is subdominant to $q_x^2$. This
implies that the WB singularity does not exist in 1D3D. This makes
sense since the bulk phonon freedom suppresses phonon fluctuations
on the chain and thus stabilizes the system against the attractive
electron-phonon coupling. Therefore, in a 1D3D system the
electron-phonon coupling at small momentums is not important no
matter the filling. A similar discussion for 1D2D leads to the same
conclusion.
%A similar calculation shows that in such systems the WB singularity
%also disappears.

\section{\label{3D3D}A 3D Array of Parallel 1D Conducting Chains Coupled to 3D phonons}
In the previous two sections we have discussed two special systems:
1D1D and 1D3D. In this section we will study a more realistic
system: a 3D array of parallel 1D conducting chains coupled to 3D
phonons.

In 1D1D we derived an exact relation between phonon and electron
correlations, which is given in (\ref{GeneralRelation}). Performing
an almost identical calculation we also find a similar result in
3D3D:
\begin{eqnarray}
 \langle\psi(\vec{q})\psi(-\vec{q})\rangle&=&{16\pi^4\alpha^2g_2^2G_{ph}^2
 (\vec{q})}\sum_{\vec{r}_{\perp}}\int dxd\tau\ \ e^{i\vec{q}\cdot\vec{r}}\times\nonumber\\
 &&\langle\rho_{2k_F}(\vec{r})\rho_{2k_F}(\vec{0})\rangle+G_{ph}(\vec{q})
 \, ,\label{GeneralRelation3D}
\end{eqnarray}
where $G_{ph}^{-1}(\vec{q})\equiv
[g_1^2+q_{\tau}^2+(g_1/2n_e)^2q_x^2+C_{\perp}q_{\perp}^2]/\pi^2$. We
have replaced the sum over chain index $j$ by the sum over
$\vec{r}_{\perp}$ with $\vec{r}_{\perp}=n\pi$ (remember that the
distance between nearest-neighbor chains is $\pi$ after the
dimensional rescaling given in (\ref{Rescaling1D3D2})), $n$ an
integer. We will adopt this change hereafter.

For a further understanding of this problem, we follow the logic
used in 1D1D and treat this 3D3D problem based on the same two
limits. Let us start with the limit of weak coupling and/or high
phonon frequency. In this limit we focus on the phonon-mediated
effective electron action:
\begin{eqnarray}
 S_{eff}/\hbar&=&{S_{el}\over \hbar}-
 %2\left(g_2^2\over \pi^{d-1}\right)
% \int d^dr_{\perp}dx dy dy'\nonumber\\
% &&\left\{\cos\left[2\phi(x, y)-
% 2\phi(x, y')\right]G'_{ph}(\vec{0},
% y-y')\right\}
 2g_2^2\sum_{\vec{r}_{\perp}, \vec{r}_{\perp}'}
 \int d^2r_{\perp} d^2r'_{\perp}dx d\tau d\tau'\nonumber\\
 &&\left\{\cos\left[2\phi(x, \tau, \vec{r}_{\perp})-
 2\phi(x, \tau', \vec{r}\, '_{\perp})\right]\times\right.
 \nonumber\\
 &&\left.G_{ph}(\tau-\tau', \vec{r}_{\perp}-\vec{r}\, '_{\perp})\right\}
 \label{InitialModel}\, ,
\end{eqnarray}
where the phonon propagator is given by
\begin{eqnarray}
 G_{ph}(\tau, \vec{r}_{\perp})
 %{1\over \left(2\pi\right)^{d+1}}\int d^dq dq_y\ \
% {S_{d-1}\over{q_y^2+C_xq_x^2+C_{\perp}q_{\perp}^2+{1\over\lambda^2_2}}}
% \ \ e^{i(\vec{q}\cdot\vec{r}+q_yy)}\nonumber\\
 &=&{1\over \left(2\pi\right)^3}\int d^2q_{\perp}dq_{\tau}\ \
 \left[{\pi^2\over{q_{\tau}^2+C_{\perp}q_{\perp}^2+g_1^2}}\right.\nonumber\\
 &&\left.\times e^{i(\vec{q}_{\perp}\cdot\vec{r}_{\perp}+q_{\tau}\tau)}\right]\, .
\end{eqnarray}
We have neglected the phonon dispersions along the chains, which
only lead to insignificant correction. The phonon dispersions
perpendicular to the chains, although typically very small, are
important since they are responsible for the existence of
long-range-order CDW. This will be seen later. Now we apply a
self-consistent Gaussian variational method \cite{Giamarchi} to this
model . We approximate the action (\ref{InitialModel}) as a
quadratic one
\begin{eqnarray}
 S_0/\hbar={1\over 2}\sum_{\vec{q}}G_{el}^{-1}(\vec{q})\phi(-\vec{q})
 \phi(\vec{q})\, .
\end{eqnarray}
Then we try to optimize the propagator $G_{el}^{-1}(\vec{q})$ so as
to minimize the variational free energy
\begin{eqnarray}
 F_v=F_0+\langle S-S_0\rangle_0\, ,
\end{eqnarray}
where both the free energy $F_0$ and the average $\langle\rangle_0$
are calculated using the quadratic action $S_0$. Performing the
procedure we obtain a self-consistent equation
\begin{eqnarray}
 G_{el}^{-1}(\vec{q})&=&{1\over \pi^3 K}(q_x^2+q_{\tau}^2)
 +16\left(g_2\over \pi\right)^2\times\nonumber\\
 &&\sum_{\vec{r}_{\perp}}\int d\tau\ \
 e^{-4\left(\langle\phi^2\rangle_0-\langle\phi(0, 0, \vec{0})\phi(0, \tau, \vec{r}_{\perp})\rangle_0\right)}
 \nonumber\\
 &&\times G_{ph}(\tau, \vec{r}_{\perp})
 \left[1-e^{i\left(\vec{q}_{\perp}\cdot\vec{r}_{\perp}+q_{\tau}\tau\right)}\right]
 \label{Self3D3}\, .
\end{eqnarray}
We notice that the right-hand side of the above equation vanishes as
$\vec{q}\to 0$. Furthermore, if we naively expand the cosine in the
action (\ref{InitialModel}) to the quadratic order of $\phi$, we
will get corrections to $|\vec{\nabla}_{\perp}\phi|^2$ and
$(\partial_{\tau}\phi)^2$ and no corrections to $(\partial_x
\phi)^2$. Therefore, we assume the following trial solution
\begin{eqnarray}
G_{el}^{-1}(\vec{q})={1\over
\pi^3K}(q_x^2+Rq_{\tau}^2+Mq_{\perp}^2)\, .\label{Trial3D3}
\end{eqnarray}
Hereafter we define $K'$ as the renormalized $K$, given by
\begin{eqnarray}
K'=K/\sqrt{R}\, .\label{DefineK'}
\end{eqnarray}
The self-consistent equation (\ref{Self3D3}) can be solved
analytically in the limit $M\ll 1$, $(K/K')^2-1\ll 1$. In this limit
the system is highly anisotropic. Electron correlations along the
chains decay as a power law at length scales shorter than
$1/\sqrt{M}$ and tend towards a non-zero constant at longer length
scales, while electron correlations along $\perp$ directions decay
very rapidly to the same non-zero constant. Specifically, they are
given as
\begin{eqnarray}
 &&e^{-\left(\langle\phi^2\rangle_0-\langle\phi(0, 0, \vec{0})\phi(0, \tau, \vec{r}_{\perp})\rangle_0\right)}
 \nonumber\\
 &=&\left\{\begin{array}{ll}
 \tau^{-{K'\over 2}}\, , &\vec{r}_{\perp}=\vec{0}\, ,\tau<1/\sqrt{M}\\
 M^{{K'\over 4}}\, , &\vec{r}_{\perp}\neq
 \vec{0}~~\mbox{or}~~\tau> 1/\sqrt{M}
 \end{array}
 \right.\, ,
 \label{ElectronCorrelation}
\end{eqnarray}
which reduce to the correlations of TLL in the limit $M\to 0$.
Plugging the formula (\ref{ElectronCorrelation}) into the equation
(\ref{Self3D3}) we obtain two coupled self-consistent equations
\begin{eqnarray}
 \left(K\over K'\right)^2-1&=&16\pi K g_2^2
 \int_{-1/\sqrt{M}}^{1/\sqrt{M}} d\tau \ \ \tau^{-{2K'}}\tau^2G_{ph}(\tau, \vec{0})
 \nonumber\\
 &&+32\pi K g_2^2
 \int_{1/\sqrt{M}}^{\infty} d\tau \ \ M^{{K'}}\tau^2G_{ph}(\tau, \vec{0})
 \nonumber\\
 &&+16\pi K M^{K'}g_2^2g_1^{-4}\, , \label{3D3R}\\
 M&=&16\pi K M^{K'}C_{\perp}g_2^2g_1^{-4}\label{3D3M}\, ,
\end{eqnarray}
where
\begin{eqnarray}
G_{ph}(\tau, \vec{0})={\pi\over 4C_{\perp}\tau}
\left[e^{-g_1\tau}-e^{-\tau\sqrt{{4C_{\perp}\over
\pi}+g_1^2}}\right]\, .
\end{eqnarray}
In the special case $C_{\perp}=0$, $M$ vanishes, and the second and
the third terms on the right-hand side of (\ref{3D3R}) thus vanish,
and then (\ref{3D3R}) reduces to the result of 1D1D. This is
expected since the 3D3D model with $C_{\perp}=0$ corresponds to a
simple sum of 1D1D models. For $C_{\perp}$ being nonzero, $M$ can be
easily calculated from (\ref{3D3M}) as
\begin{eqnarray}
M&=&\left\{\begin{array}{ll}0\, , &K'>1\, ,\\
(16\pi KC_{\perp}g_2^2g_1^{-4})^{1\over {1-K'}}\, ,&K'<1\, .
\end{array}
\right. \label{Self3D3M}
\end{eqnarray}
After a more involved, but essentially straightforward calculation,
%Since the form of
%$G_{ph}(\tau, \vec{0})$ depends on whether $C_{\perp}\ll g_1^2$ or
%$C_{\perp}\gg g_1^2$, we need to discuss the two cases separately.
%We need to consider the complications whether $M\ll g_1^2$ or $M\gg
%g_1^2$, since the second term on the right side of (\ref{3D3R}) will
%vanish in the former case and dominate over the first term in the
%latter. After a but tedious calculation
we get, to the lowest order of $(K/K')^2-1$,
\begin{eqnarray}
\left(K\over K'\right)^2-1&\sim& g_2^2g_1^{2K-4}
\label{SmallSelf3D3R}
\end{eqnarray}
for $C_{\perp}\ll g_1^2$ and
\begin{eqnarray}
\left(K\over K'\right)^2-1&\sim& \left\{\begin{array}{ll}
{1\over C_{\perp}}\left[M+g_2^2g_1^{2K-2}\right]\, , &M\ll g_1^2\\
{M/C_{\perp}}\, , &M\gg g_1^2
\end{array}
\right. \label{BigSelf3D3R}
\end{eqnarray}
for $C_{\perp}\gg g_1^2$. The solution of $M$ implies a quantum
phase transition at about $K=1$ for arbitrary weak electron-phonon
coupling. This critical point can also be obtained by calculating
the average of the retarded part of the action for a single chain
with respected to the free electron action. This average scales as
$L^{2-2K}$ and thus also predicts the critical point $K=1$. For
$K<1$, electrons on different chains are effectively correlated,
which leads to long-range-order CDW. The order parameter is the
$2k_F$-wave component of the electron density, whose average is {\it
non-zero} and scales as
\begin{eqnarray}
 \langle\rho_{2k_F}\rangle\sim{M^{K'\over 2}\over {2\pi\alpha}}\,
 .\label{ExpectationCharge}
\end{eqnarray}
%At finite temperature the model becomes a
%classical quasi-one dimensional system with weak inter-chain
%coupling \cite{ClassicChain}.
According to the general result given in (\ref{GeneralRelation3D}),
the $2k_F$-wave component of the phonon displacement field also
becomes long-range-ordered with its expectation value
\begin{eqnarray}
 \langle\psi\rangle\sim {g_2\over g_1^2}M^{K'\over 2}\,
 .\label{ExpectationPhonon}
\end{eqnarray}
For $K>1$, electrons on different chains are effectively not
correlated, and each chain behaves as an isolated single 1D
conducting chain coupled to 3D phonons, which has been discussed in
section \ref{1D3D}. For self-consistency, the solutions
(\ref{Self3D3M}), (\ref{SmallSelf3D3R}) and (\ref{BigSelf3D3R}) have
to satisfy $M\ll 1$, $(K/K')^2-1\ll 1$, which leads to a restriction
on the parameters
\begin{eqnarray}
 \left\{\begin{array}{ll}
 {g_2\over g_1}\ll g_1^{1-K}\, ,&C_{\perp}\ll g_1^2\, ,\\
 {g_2\over g_1}\ll {g_1\over\sqrt{C_{\perp}^K}}\, ,&C_{\perp}\gg g_1^2\, .
 \end{array}\right.
 \label{WeakCondition3D}
\end{eqnarray}
This condition precludes the application of the Gaussian variational
result in the region above $OA$ in figure \ref{fig: StrongWeak}.

%\subsection{\label{LS}Low Phonon Frequency and/or Strong Coupling}

For stronger coupling and/or lower phonon frequency, the result we
get from the Gaussian variational method is not valid since the
condition (\ref{WeakCondition3D}) is violated. In this case we have
to use the fixed-amplitude approximation, which has been discussed
in section \ref{1D1D}. We assume that the important fluctuations of
$\psi$ are its phase fluctuations, and the amplitude of $\psi$ can
be approximated as fixed at its saturated value $\psi_0$. After this
approximation we obtain a simplified model
\begin{eqnarray}
 S_{total}^{3D}/\hbar&=&
 {1\over 2\pi^3}\int dxd^2r_{\perp} d\tau \left\{{1\over K}
 \left[(\partial_{\tau}\phi)^2 +
 (\partial_x \phi)^2\right]+\right.\nonumber\\
 &&\left.\left[(\partial_{\tau}\theta)^2
 +g_1^2\left(1\over 2n_e\right)^2
 (\partial_x \theta)^2+C_{\perp}|\vec{\nabla}_{\perp}\theta|^2
 \right]\right.\nonumber\\
 &~&\left.\times\pi\psi_0^2\right\}-{2g_2\psi_0\over \pi^2}\int dxd^2r_{\perp} d\tau \cos\left(2\phi-\theta\right)
 \label{Total3D3}\, ,\nonumber\\
\end{eqnarray}
where we have transformed the action into its continuum version, and
the constant pieces have been thrown away. We will calculate the
fluctuations of $\phi$ and $\theta$ based on this truncated model.

The locally saturated value of $\psi$ is also estimated by using the
mean-field theory, which approximates $\psi$ as uniform both in
space and time. Since there are no bare electron correlations
between different chains, the mean-field 3D3D model is a simple sum
of the mean-field 1D1D models. Therefore, the mean-field calculation
is the same as that in 1D1D, and the result is thus also the same;
that is, $\psi_0$ is nonzero and given by the formula (\ref{Psi})
for $K<1$, and zero for $K>1$. However, unlike in 1D1D, $\psi$ is
now {\it truly} long-range-ordered for $K<1$. In the following we
will include the phase fluctuations and calculate the non-zero
expectation value of $\psi$.
Now we discuss the truncated model (\ref{Total3D3}). It can be shown that the cosine term in the
model is always relevant no matter the value of $\psi_0$,
as long as $K$ is less than 2. The proof is given in appendix \ref{PsiC}.
%As in 1D1D, for the model (\ref{Total3D3}) there also exists a
%critical value of $\psi_0$. For $\psi_0$ bigger than this critical
%value, the cosine term in the model (\ref{Total3D3}) is relevant,
%and
Therefore, for $K<1$ the fluctuations of $\phi$ and $\theta$ are
always bound together,
and both are characterized by the same action
\begin{eqnarray}
 S[\Xi]/\hbar&=&{1\over 2\pi^3 K}\int dxd^2r_{\perp}d\tau\left[R(\partial_{\tau}\Xi)^2
 +B(\partial_x\Xi)^2+\right.\nonumber\\
 &&\left.M|\vec{\nabla}_{\perp}\Xi|^2\right]\, ,
 \label{Fukuyama2}
\end{eqnarray}
where $\Xi$ can be either $\phi$ or $\theta/2$, parameters $R$, $B$ and $M$ are given
by
\begin{eqnarray}
 R&=&(1+4\pi K\psi_0^2)\label{Strong3DR}\, ,\\
 B&=&1+4\pi Kg_1^2\left(1\over 2n_e\right)^2\psi_0^2
 \, ,\\
 M&=&C_{\perp}\pi\psi_0^2\, .\label{StrongM}
\end{eqnarray}
Note that the correction to $B$ is negligible compared to the
correction to $R$, since $g_1$ is much less than 1. Plugging the
expression of $\psi_0$ into (\ref{Strong3DR}) and (\ref{StrongM})
and writing $R$ in terms of $K'$ using (\ref{DefineK'}), we get
\begin{eqnarray}
 &&\left(K\over K'\right)^2-1\sim (g_2^2g_1^{2K-4})^{1\over (1-K)}\, ,\\
 &&M\sim C_{\perp}(g_2^2g_1^{2K-4})^{1\over (1-K)}\, ,
\end{eqnarray}
which agrees with the result in reference \cite{Artemenko}. This
result is consistent with the one obtained by the Gaussian
variational method in that both predict long-range-ordered CDW and
hence long-range-ordered $\psi$ for $K<1$. The expectation values of
$\rho_{2k_F}$ and $\psi$ are {\it non-zero} and given by the
formulae (\ref{ExpectationCharge}) and (\ref{ExpectationPhonon}),
respectively. Clearly this fixed-amplitude approximation can {\it
not} give result for attractive electron-electron interactions since
the method relies on the condensation of the phonon field $\psi$,
which is only possible for repulsive electron-electron interactions
according to the mean-field theory. Based on the result of weak
coupling and/or high phonon frequency, we suspect that for strong
coupling and/or low phonon frequency, there also exists a critical
$K$ above which both electron and phonon correlations become quasi
long-ranged; however, this critical $K$ could be significantly
bigger than 1.
%The latter scales as
%\begin{eqnarray}
%  \langle\psi\rangle&\sim&(g_2^2g_1^{2K-4})^{1\over {1-K}}M^{K'\over
%  2}\, ,
% \end{eqnarray}
%which is  and thus implies that $\psi$ is also long-range-ordered.

Now let us discuss the validity of the fixed-amplitude
approximation. Unlike in 1D1D, based on the above discussions
there is no self-consistent condition cause by phase fluctuations in 3D3D.
However, the one caused by amplitude fluctuations still exists. After
a similar calculation as we did in 1D1D, we find that the fixed-amplitude
approximation is self-consistent {\it only} if
%For self-consistency, the fixed-amplitude
%approximation is good {\it only} if
%\begin{eqnarray}
% \langle(\delta\psi_0)^2\rangle\ll \psi_0^2\, .
% \label{AmplitudeCondition}
%\end{eqnarray}
%For a crude estimation of $\langle(\delta\psi_0)^2\rangle$, we
%expand the action around the saddle point by writing
%$\psi=(\psi_0+\delta\psi_0)e^{i\theta}$, and assume that the
%effective action for $\delta\psi_0$ has the following form
%\begin{eqnarray}
% S[\delta \psi_0/\hbar&=&{1\over 2\pi^2}\int dx d\tau\ \
% \left[{{2(1-K)}g_1^2\over {2-K}}(\delta \psi_0)^2 +
% (\partial_{\tau} \delta\psi_0)^2+C_{\perp}(\nabla_{\perp}\delta\psi_0)^2\right.\nonumber \\
% &&\left.+g_1^2\left(1\over
% 2n_e\right)^2(\partial_x\delta\psi_0)^2
% \right]
% \, ,
% \label{ExpansionAroundSaddlePoint}
%\end{eqnarray}
%where the mass term is expected when expanding around a minimum, and
%the derivative terms come from the original phonon action. Then
%after a straightforward calculation, condition
%(\ref{AmplitudeCondition}) leads to
\begin{eqnarray}
 \left\{\begin{array}{ll}
 {g_2\over g_1}\gg g_1^{1-K\over 2}\, ,&C_{\perp}\ll g_1^2\, ,\\
 {g_2\over g_1}\gg g_1^{1-K\over 2}\left(g_1\over\sqrt{C_{\perp}}\right)^{1-K\over
 2}\, ,&C_{\perp}\gg g_1^2\, ,
 \end{array}
 \right.\label{StrongCondition3D}
\end{eqnarray}
which prohibits the application of the fixed-amplitude approximation
in the region below $OB$ in figure \ref{fig: StrongWeak}.
%This is consistent with the result from the limit of weak coupling
%and/or high phonon frequency.

According to the self-consistent conditions (\ref{WeakCondition3D})
and (\ref{StrongCondition3D}), there is an intermediate region in
the parameter space of $g_1$ and $g_2$ in which neither of the
approaches applies. This region is illustrated in figure \ref{fig:
StrongWeak} as the area between the loci $OB$ and $OA$. In this
region, we can not obtain a quantitatively trustable result;
however, we do expect long-ranged electron and phonon correlations
for repulsive electron-electron interactions. We also want to point
out that
for $C_{\perp}\sim g_1^2$, both approaches give the same result:
%in the whole parameter space of $g_1$ and $g_2$ both
%results reduce to
\begin{eqnarray}
 M&\sim& \left(g_2\over g_1\right)^{2\over {1-K}}\, ,\\
 \left(K\over K'\right)^2-1&\sim&(g_2^2g_1^{2K-4})^{1\over {1-K}}\,
 .
\end{eqnarray}
This strongly suggests that in this special case, this quantitative
result should also hold in the intermediate region.

So far we have focused on the electron-phonon coupling at large
momentums. In section \ref{1D3D} we discussed the effect of the
electron-phonon coupling at small momentums in 1D1D and 1D3D.
Repeating the (virtually same) calculation, we find that the WB
singularity also exists in 3D3D and is given by the formula
(\ref{WB}). Again, for a general filling, according to the
postulates (\ref{BasicAssumption}) the system is far away from the
WB singularity; therefore, the electron-phonon coupling at small
momentums is not important.

\section{Conclusion}
%In the limit of
%weak coupling and/or low phonon frequency, we use a Gaussian
%variational method,  treats the phonon fluctuations {\it fully} and
%gives satisfying result for electrons; however, it can {\it only} be
%solved analytically for weak coupling and/or high phonon frequency,
%and  On the other hand, while the fixed-amplitude approximation
%allows us to obtain the result for
%the
%strong-coupling approach is able to calculate
%both electrons and phonons,
%the expectation values of $\rho_{2k_F}$
%and phonon field $\psi$; however,
%it is {\it only} self-consistent for strong coupling and/or low
%phonon frequency.
In this paper we have studied electron and phonon correlations in
systems of one-dimensional spinless electrons coupled to phonons at
low temperatures. The focus is on the effect of backward
electron-phonon scattering. We have been able to obtain quantitative
results in the limits of weak coupling and/or high phonon frequency
and strong coupling and/or low phonon frequency, which then leads to
a qualitative understanding in the intermediate region as well.

It has to be mentioned that our work for a three-dimensional array
of one-dimensional conducting chains coupled to three-dimensional
phonons does not take into account the effect of the interchain
hopping. It is well known that the interchain hopping can break down
the TLL behavior which we have assumed for electrons in our
calculations. On the other hand, it is not clear if the effect of
the interchain hopping will be suppressed in the presence of
electron-phonon coupling. To understand these questions, a work
which includes both effects of electron-phonon coupling and the
interchain hopping will be interesting.

\section*{Acknowledgment}
The author is grateful to John Toner, Thomas Nattermann and Zoran
Ristivojevic for valuable discussions. The author acknowledges
support from the SFB 608.

\appendix
\section{\label{HW1D1D}Weak Coupling and/or High Phonon Frequency in 1D1D}
Since the total action is Gaussian in the phonon field $\psi$, we
integrate out the phonon freedom and get an effective action only
involving the electron field $\phi$
\begin{eqnarray}
 S_{el}^{1D}/\hbar&=&{1\over 2\pi K}\int dx d\tau\ \
 \left[(\partial_{\tau}\phi)^2 + (\partial_x \phi)^2\right]
 \nonumber\\
 &~&-{g_2^2\over g_1}
 \int dx d\tau d\tau'\ \ G_{ph}(\tau-\tau')
 \cos\left[2\phi(x, \tau)-\right.\nonumber\\
 &&\left.2\phi(x, \tau')\right]\, .
 \label{AEffectElectron1D1}
\end{eqnarray}
where the phonon correlation function is given by
\begin{eqnarray}
 G_{ph}(\tau)
 \approx e^{-g_1|\tau|}\, .
\end{eqnarray}
We have neglect the phonon dispersions $(\partial_x\phi)^2$ since
their effect is not important. When we discuss an array of
paralleled 1D chains coupled to 3D phonons, we keep the phonon
dispersions perpendicular to the wires since they can lead to
long-range-order CDW.

First we apply a Gaussian variational calculation to the phonon
mediated effective action (\ref{AEffectElectron1D1}). We assume a
Gaussian variational electron action
\begin{eqnarray}
 S_0=\sum_{\vec{q}}G_{el}^{-1}(\vec{q})\phi^*(\vec{q})
 \phi(\vec{q})\,
 \label{Gaussian1}
\end{eqnarray}
and try to optimize it to minimize the variational free energy
\begin{eqnarray}
 F_{var}=F_0+\langle S_{el}^{1D}-S_0\rangle_0\, ,
\end{eqnarray}
where $F_0$ and $\langle\rangle_0$ are calculated using action
(\ref{Gaussian1}). After a straightforward calculation, we obtain
\begin{eqnarray}
 G_{el}^{-1}(q_x, q_{\tau})
 %&=&{1\over \pi K}(q_x^2+q_y^2)
% +8\left(1\over\lambda_2^2\right)^2
% \int dy\ \ \lambda_1e^{-y/\lambda_1}
% e^{-4\left[G_{el}(0, 0)-G_{el}(0, y)\right]}
% \nonumber\\
% &~&\left[1-e^{iq_yy}\right]
 &=&{1\over \pi K}(q_x^2+q_{\tau}^2)
 +{8g_2^2\over g_1}
 \int_{-\infty}^{+\infty} d\tau\ \ e^{-g_1|\tau|}\times \nonumber\\
 &&e^{-4\left[G_{el}(0)-G_{el}(\tau)\right]}
 \left[1-\cos(q_{\tau}\tau)\right]\label{Self7}
 \, ,
\end{eqnarray}
where
\begin{eqnarray}
 G_{el}(\tau)={1\over 4\pi^2}\int dq_xdq_{\tau}\ \
 G_{el}(\vec{q})e^{iq_{\tau}\tau}\, .
\end{eqnarray}
Since the right-hand side of the above equation vanishes as
$\vec{q}\to \vec{0}$, and the integral is independent of $q_x$, we
assume the following trial solution
\begin{eqnarray}
 G_{el}^{-1}(q_x, q_{\tau})={1\over \pi K'}
 \left({1\over\sqrt{R}}q_x^2+\sqrt{R}q_{\tau}^2\right)
 \label{ElectronPropagator}
\end{eqnarray}
with $K'\equiv K/\sqrt{R}$. With the electron propagator written in
this way, we ensure that the coefficient of $q_x^2$ is not
renormalized but fixed at $1/(\pi K)$. Hereafter we denote $K'$ and
$K$ as the renormalized and initial value respectively. Then the
electron correlation function can be calculated as
\begin{eqnarray}
 G_{el}(0)-G_{el}(\tau)%&=&{1\over 4\pi^2}\int dq_x dq_y\ \
% {\pi K\over (q_x^2+Rq_y^2)}\left[1-\right.\nonumber\\
% &&\left.e^{iq_yy}\right]\nonumber\\
 %&=&{K\over 2\sqrt{R}}\ln{\left(Ry+a\over a\right)}
 ={K'\over 2}\ln{\left(K^2\alpha \tau+aK'\, ^2\over a K'\, ^2\right)}\, .
\end{eqnarray}
Plugging this result into Eq. (\ref{Self7}), we obtain
\begin{eqnarray}
 &&{1\over \pi K}\left[\left({K^2\over K'\, ^2}-1\right)q_{\tau}^2\right]\nonumber\\
 %&=&{8\lambda_1\over\lambda_2^4}
% \int dy \ \ e^{-y/\lambda_1}
% \left(Ry+a\over a\right)^{-{2K\over\sqrt{R}}}
% \left[1-e^{iq_yy}\right]
 &=&{8g_2^2\over g_1}
 \int_{-\infty}^{+\infty} d\tau \ \ e^{-g_1|\tau|}
 \left(K^2\alpha \tau+aK'\, ^2 \over aK'\, ^2 \right)^{-2K'}\nonumber\\
 &&\times\left[1-\cos(q_{\tau}\tau)\right]
 \, .
\end{eqnarray}
For $q_{\tau}\to 0$, this leads to
\begin{eqnarray}
 &&{1\over \pi K}\left({K^2\over K'\, ^2}-1\right)\nonumber\\
 %&=&{8\lambda_1\over\lambda_2^4}
% \int dy \ \ e^{-y/\lambda_1}
% \left(Ry+a\over a\right)^{-{2K\over\sqrt{R}}}
% \left[1-e^{iq_yy}\right]
 &=&8\left(g_2^2\over g_1\right)
 \int_{\alpha/a}^{+\infty} d\tau \ \ \tau^2e^{-g_1|\tau|}
 \left(\alpha K^2\tau\over a K'\, ^2\right)^{-2K'}
 \, .
 \nonumber\\
 \label{SelfK'}
\end{eqnarray}
It is interesting to point out that in the absence of
$e^{-g_1|\tau|}$ in the integrant on the right-hand side of
(\ref{SelfK'}), the integral would converge for $K'>3/2$ and diverge
for $K'<3/2$, and there would be a quantum phase transition at the
critical point $K'=3/ 2$. The two phases are characterized by $K'\to
0$ for $K'<3/2$ and finite $K'$ for $K'>3/2$ respectively. With the
presence of $e^{-g_1|\tau|}$, the integral in (\ref{SelfK'})
converges for any $K'$. Thus, there is no phase transition. Assuming
$K'$ is close to $K$, calculating $K'$ to the lowest-order
correction $K'-K$ we get
%\begin{eqnarray}
% I=2(\pi)^{2K}g_1^{2K-3}\Gamma(3-2K)\, ,
% \label{I}
%\end{eqnarray}
\begin{eqnarray}
 {K\over K'}-1\sim \left\{\begin{array}{ll}
 -\left(g_2^2\over g_1\right)
 g_1^{2K-3}\, &K<{3\over 2}\\
 {g_2^2\over g_1}\ln{g_1}\, &K={3\over 2}\\
 -{g_2^2\over g_1}\, &K>{3\over 2}
 \end{array}\right.\, .
 \label{AK1}
\end{eqnarray}
%From this result we see that $K-K'$ diverges for $K<{3/2}$ if we let
%$g_1\to 0$ while keep $g_2^2/g_1$ fixed, which indicates the phase
%transition we just discussed.
Although there is no phase transition,  there exists a sharp
crossover at $K=K_s$ for $K<3/2$ with
\begin{eqnarray}
 K_s={3\over 2}-{\ln\left(g_2^2\over g_1\right)\over 2\ln{g_1}}\, ,
 \label{Ks}
\end{eqnarray}
which is obtained by doing $(K/K')-1=1$. The effect of
electron-phonon coupling on electrons is significant for $K<K_s$ and
negligible for $K>K_s$. This can be easily seen if we examine the
renormalized plasma velocity $v'=v/\sqrt{R}$ where $R$ is given by
\begin{eqnarray}
 R-1\sim
 %\left\{\begin{array}{ll}
 \left(g_2^2\over g_1\right)
 g_1^{2K-3}\, .
% &K<{3\over 2}\\
% -{g_2^2\over g_1}\ln{g_1}\, &K={3\over 2}\\
% {g_2^2\over g_1}\, &K>{3\over 2}
% \end{array}\right.\, .
 \label{R}
\end{eqnarray}
Since both $g_2^2/g_1$ and $g_1$ are typically much less than 1, for
$K>K_s$ $R$ is always close to 1, and $v'$ gets hardly renormalized.
For $K\to K_s^+$, $R$ suddenly becomes larger than 1, (Strictly
speaking, formula (\ref{R}) doesn't apply anymore for $K<K_s$, but
it qualitatively tells the trends nonetheless.) and $v'$ is thus
reduced appreciably. In principle this result can be tested by
experiments since $K_s$ can be less than 1 according to (\ref{Ks}).
%\begin{eqnarray}
% R=8\pi K(\pi)^{2K}g_2^2g_1^{2K-4}+1\, .
% \label{R1}
%\end{eqnarray}

Now we apply the standard momentum shell renormalization group
calculation (RG) to this model. We separate the electron field into
fast and slow varying components $\phi=\phi^>+\phi^<$, where
$\phi^>$ has support in the momentum shell
$ae^{d\ell}/\alpha<q_x<a/\alpha$, $-\infty<q_{\tau}<\infty$,
integrate out the fast varying component $\phi^>$ and rescale the
length with $x=x'e^{d\ell}$, $\tau=\tau'e^{d\ell}$ so as to restore
the original ultraviolet cutoff. The integration over $\phi^>$ is
achieved perturbatively by expanding the partition function to the
first order around the Gaussian fixed point. Performing the above
procedure and using the expression (\ref{ElectronPropagator}) for
the propagator, we obtain the following RG flow equations (up to one
loop)
\begin{eqnarray}
 {d\left[K'(\ell)\right]\over d\ell}&=&-C_1K'\,^3{g_2^2\over g_1}(\ell)f(g_1(\ell))\, ,\label{K'}\\
 %{d\left[R(\ell)\right]\over d\ell}&=&2C_1K^2{g_2^2\over g_1}(\ell)\,
% ,\label{R}\\
 {d\left[{g_2^2\over g_1}(\ell)\right]\over d\ell}&=&\left(3-2K'\right)
 {g_2^2\over g_1}(\ell)\, ,\label{Bg2}\\
 {d\left[g_1(\ell)\right]\over d\ell} &=& g_1(\ell)\, ,\label{g1}
\end{eqnarray}
where $C_1$ is a constant of order 1, $f(x)$ is a step function
which is 1 for $x<1$ and 0 for $x>1$. We use $f(x)$ to cut the
renormalization of $K'(\ell)$ at
$\ell_1=-\ln\left(g_1a/\alpha\right)$, where the phonon propagator
$G_{ph}(\tau-\tau')$ becomes effectively $\delta(\tau-\tau')$. The
correspondence between our notation and that of reference
\cite{Void} is: $Y_{ph}\equiv g_1/g_2$, $g_1\sim\alpha/\xi_{ph}$,
$exp(2\xi)\equiv K'$. It seems that the correct eigenvalue in the RG
flow equation (3.14b) in reference \cite{Void} should be
$2-2exp(2\xi)$. Also there are some minor differences between our RG
flow equations and those in reference \cite{Void}. They are due to
different RG procedures and should not affect the physics. The RG
flow Eq. (\ref{Bg2}) seems to imply a phase transition at $K'=3/2$;
however, since the renormalization of $K'(\ell)$ is utterly cut at
$\ell=\ell_1$, $K'$ is always finite even for $K'<3/2$, so there
will be no phase transition at $K'=3/2$.
%Furthermore, for $K>K_s$,
%$g_2^2(\ell)/g_1(\ell)$ remains less than 1 at $\ell=\ell_1$; for
%$K>K_s$, the RG flows to the strong-coupling limit at
%$\ell=\ell_2<\ell_1$ with
%$\ell_2=\ln\left[g_1(\ell)/g_2^2(\ell)\right]/(3-2K)$. This implies
%the sharp change at $K=K_s$ we mentioned earlier.
For $g_2g_1^{K-2}<1$, the perturbative RG holds way down to
$\ell=\ell_1$ since $g_2^2(\ell_1)/g_1(\ell_1)$ remains less than 1,
and
%For $K>1$, condition (\ref{Condition1}) is always satisfied since
%$g_2(\ell)/g_1(\ell)$ decrease under the rescaling and the above RG
%flow equations hold all the way down to $\ell^*\sim \ln{g_1}$ after
%which the renormalization of $R$ is negligible since $g_1(\ell)$
%reaches the cutoff $a/\alpha$. Therefore
we can integrate the flow equations approximately to calculate $K'$
as
\begin{eqnarray}
(K'-K)&\sim&-{g_2^2\over g_1}\int_0^{\ell^*} e^{(3-2K)\ell} d\ell\,
,
    %&=&-{g_2^2\over g_1}\int_0^{\ell^*} d\ell\ \
%    e^{(3-2K)\ell}\nonumber\\
    %&\sim&{-1\over {2K-3}}\left(g_2^2\over g_1\right)\left[g_1^{2K-3}-
%    \left(\alpha\over a\right)^{3-2K}\right]\, ,
\end{eqnarray}
which essentially gives the Gaussian variational result (\ref{AK1}).
For $g_2g_1^{K-2}>1$, $g_2^2(\ell)/g_1(\ell)$ becomes larger than 1
at $\ell=\ln\left[g_1(\ell)/g_2^2(\ell)\right]/(3-2K)$, which is
less than $\ell_1$. In this case the perturbative RG breaks down
before we can calculate $K'$; therefore, a different strategy needs
to be used to attack the problem.

\section{\label{1D1DStrongLow}Calculation of $\psi_0$}
%In this limit, the perturbative RG breaks down. A different strategy
%has been developed \cite{Void}. For strong coupling or low phonon
%frequency, it is expected that the magnitude of the $2k_F$ component
%of the phonon field sutures locally at some big value such that its
%fluctuation can be ignored. After realizing this we can simplify the
%model by writing $\psi=\psi_0e^{i\theta}$ with $\psi_0$
%non-fluctuating. Then the simplified model can be easily solved.
%In this section we describe a different approach to calculate
%$\psi_0$. Compared to the approach used in ref. \cite{Void}, our
%approach is less accurate but simple, and it gives qualitatively
%correct result for $K$ not very close to 1. The calculation also
%starts with the mean-field approximation that $\psi$ is uniform both
%in space and time.
After the mean-field approximation we get a simplified model
%First Let's discuss how to estimate the magnitude $\psi_0$. Previous
%work \cite{Void} did this calculation using the mean-field theory.
%At the first step field $\psi$ is approximated as uniform both in
%space and time. At the second step the model is mapped to an exactly
%solvable field-theoretical model (i.e., Massive Thirring model) and
%the ground state energy of the electron is calculated in terms of
%$\psi_0$. This leads to an effective Hamiltonian for $\psi_0$. At
%the third step this Hamiltonian is minimized with respect to
%$\psi_0$ to give non-zero $\psi_0$ for $K<1$ and zero $\psi_0$ for
%$K>1$.  In this paper we follow logically the same procedure but try
%a different approach at the second step. We integrate out the
%electron freedom in favor of $\psi_0$ brutally to get an effective
%action for $\psi_0$. The calculation is not done exactly but with a
%Gaussian variational method. We obtain qualitatively the same result
%as that in \cite{Void}. However, our approach doesn't recover the
%classical Peierls theory at $K=1$ due to the approximation we use.
%Apparently the strategy discussed in this section relies on an
%non-zero $\psi_0$ and thus only applies for $K<1$.
%After approximating field $\psi$ as unform $\psi(x, \tau)\equiv
%\psi_0$, we obtain
%\begin{eqnarray}
% S&=&S_{el}^{1D}+S_{int}^{1D}+S_{ph}^{1D}
%\begin{eqnarray}
%with
\begin{eqnarray}
 {S_{el}^{1D}\over\hbar}+{S_{ep}^{1D}\over\hbar}
 &=&{1\over 2}\int dx d\tau \ \ {1\over\pi K}
 \left[(\partial_{\tau}\phi)^2 +
 (\partial_x \phi)^2\right]\nonumber\\
 &&-2g_2\psi_0\int dx d\tau\ \ \cos\left(2\phi\right)\, ,
 \label{SineGordon}\\
 S_{ph}^{1D}/\hbar&=&{1\over 2}\int dx d\tau\ \ g_1^2\psi_0^2\, .
\end{eqnarray}
%Now
%we try to eliminate the electron freedom. In ref. \cite{Void}
%this is done by mapping the model to the Massive Thirring model and
%calculating the ground state energy of electrons.
%we integrate out the electron field approximately using a Gaussian
%variational method.
The partition function can be calculated as
\begin{eqnarray}
 Z&=&\int D[\psi_0]D[\phi]\ \ e^{-{1\over\hbar}(S_{el}^{1D}+S_{ph}^{1D}+S_{ep}^{1D})}\nonumber\\
 &=&\int D[\psi_0]\ \ e^{-S_{ph}^{1D}/\hbar} \int D[\phi]\ \ e^{-{1\over\hbar}(S_{el}^{1D}+S_{ep}^{1D})}\nonumber\\
 &\equiv&\int D[\psi_0]\ \ e^{-{1\over\hbar}(S_{ph}^{1D}+F_{el}[\psi_0])}
 \label{Partition1}\, ,
\end{eqnarray}
where $F_{el}[\psi_0]$ is the free energy of electrons for action
$S_{el}^{1D}+S_{ep}^{1D}$ at a fixed $\psi_0$.
$S_{el}^{1D}+S_{ep}^{1D}$ is exactly a Sine-Gordon model. Formula
(\ref{Partition1}) defines an effective action for $\psi_0$
\begin{eqnarray}
 S_{ph}^{eff}[\psi_0]=S_{ph}^{1D}+F_{el}[\psi_0]\, .
 \label{Effective}
\end{eqnarray}
To calculate $F_{el}[\psi_0]$, we use Gaussian variational method
\cite{Giamarchi}. For any action $S_0$, we have
%We write the partition function for this
%model as the following
%\begin{eqnarray}
% Z&=&\int D\phi\ \ e^{-S}\nonumber\\
% &=&\int D\phi\ \ e^{-S_0}e^{-(S-S_0)}\nonumber\\
% &=&Z_0\left<e^{-(S-S_0)}\right>_0
%\end{eqnarray}
%where the index $0$ denote the partition function and the average
%with respect to $S_0$ which can be any action. Thus the free energy
%freedom satisfies
\begin{eqnarray}
 F_{el}[\psi_0]/\hbar={F_0\over\hbar}-\ln{\langle e^{-{1\over\hbar}(S_{el}^{1D}+S_{ep}^{1D}-S_0)}\rangle_0}\, .
\end{eqnarray}
where the $\langle\rangle_o$ denotes the average using the action
$S_0$. Given the convexity of the exponential \cite{Feynman} one has
always
\begin{eqnarray}
 \langle e^{-{1\over\hbar}(S_{el}^{1D}+S_{ep}^{1D}-S_0)}\rangle > e^{-{1\over\hbar}\langle S_{el}
 ^{1D}+S_{ep}^{1D}-S_0\rangle}\,
\end{eqnarray}
and thus
\begin{eqnarray}
 F_{el}^{1D}<F_{var}=F_0+\left<S-S_0\right>_0\, .
 \label{VarFree1}
\end{eqnarray}
The basic idea is to optimize $S_0$ such that $F_{var}$ gets to
$F_{el}^{1D}$ as close as possible. Assuming that $S_0$ is Gaussian
\begin{eqnarray}
 S_0/\hbar={1\over 2}
     \sum_{\vec{q}}\phi(-\vec{q})G^{-1}(\vec{q})\phi(\vec{q})\,
     \label{TrialAction}
\end{eqnarray}
with
\begin{eqnarray}
 G^{-1}(\vec{q})={{q^2+D^2}\over \pi K}\, ,
\end{eqnarray}
$F_{var}$ can be calculated as
\begin{eqnarray}
 F_{var}%&=&-\sum_{\vec{q}>0}\ln{G(\vec{q})}+{1\over 2\pi
%           K}\sum_{\vec{q}}q^2G(\vec{q})-\nonumber\\
%         &&{2g_2\psi_0}
%           \int dx dy \left<\cos{2\phi}\right>_0\nonumber\\
        &=&-\sum_{\vec{q}>0}\ln{G(\vec{q})}+{1\over 2\pi
           K}\sum_{\vec{q}}q^2G(\vec{q})-\nonumber\\
         &&{2g_2\psi_0}
           \int dx d\tau\ \ e^{-2\left<\phi^2\right>_0}\, .
 \label{VarFree2}
\end{eqnarray}
Minimizing $F_{var}$ with respect to $G(\vec{q})$ by doing $\partial
F_{var}/\partial G(\vec{q})=0$ leads to a self-consistent equation
for $D$
\begin{eqnarray}
 D^2={8\pi Kg_2\phi_0}
    \left[D\over{1+\sqrt{1+D^2}}\right]^K\, .
    \label{SelfD}
\end{eqnarray}
%\begin{eqnarray}
% G^{-1}(\vec{q})={{q^2+D^2}\over \pi K}
%\end{eqnarray}
%with
%\begin{eqnarray}
% D^2%&=&{8\pi Kg_2\phi_0}e^{-2\left<\phi^2\right>}\nonumber\\
%    %&=&{8\pi Kg_2\phi_0}
%%    \left[D\over{1+\sqrt{1+D^2}}\right]^K\nonumber\\
%    ={8\pi Kg_2\psi_0}\left(D\over 2\right)^K
%    %\left[1-K
%%    \left(D\over 2\right)^2\right]
%    \, .
%    \label{Self1}
%\end{eqnarray}
Assuming that $g_2\psi_0$ is much less than 1 (i.e., the single
electron gap is much less than the Fermi energy), we get
%\begin{eqnarray}
% D=0\label{Self3}\, ,
%\end{eqnarray}
%for $K<2$, it has a non-zero solution
\begin{eqnarray}
 D=2\left(2\pi Kg_2\psi_0\right)^{1/(2-K)}
 %-{K\over 2-K}\left(2\pi
% Kg_2\psi_0\right)^{3/(2-K)}
 \,
 \label{Self2}
\end{eqnarray}
for $K<2$ and $D=0$ for $K>2$. Therefore, $K=2$ is a naive
estimation of the critical value of $K$ at which the KBT transition
happens for the Sine-Gordon model (\ref{SineGordon}). Plugging
(\ref{Self2}) into the formula (\ref{VarFree2}), we get the
approximated $F_{el}^{1D}$
\begin{eqnarray}
 F_{el}^{1D}[\psi_0]/\hbar&=&{1\over 2\pi K}\int dx d\tau
 \left[-(2-K)\left(2\pi Kg_2\psi_0\right)^
 {2\over {2-K}}\right]\, ,\nonumber\\
\end{eqnarray}
where we have thrown out terms which are independent of $\psi_0$.
Thus, according to (\ref{Effective}) the effective action for
$\psi_0$ is given by
\begin{eqnarray}
 S_{ph}^{eff}[\psi_0]/\hbar&=&{1\over 2\pi K}\int dx d\tau
 \left[-(2-K)\left(2\pi Kg_2\psi_0\right)^
 {2\over {2-K}}+\right.\nonumber\\
 &&\left.\pi Kg_1^2\psi_0^2\right]
 %S_{ph}^{1D}[\psi_0]&=&{1\over 2\pi K}\int dx dy
% \left[-(2-K)\left(2\pi Kg_2\phi_0\right)^
% {2\over {2-K}}
% +K\left(2\pi Kg_2\phi_0\right)^{4\over{2-K}}
% \right]\nonumber\\
% &~&+{1\over 2}\int dx dy\ \ g_1^2\psi_0^2\, .
 \label{DeltaPhonon1}\, .
\end{eqnarray}
Minimize it with respect to $\psi_0$, we get $\psi_0=0$ for $K>1$
and
\begin{eqnarray}
 \psi_0=g_1^{-1}\left[2(\pi K)^{K\over 2}\left(g_2\over g_1\right)
 \right]^{1\over 1-K}
 \label{ASolution1}\,
\end{eqnarray}
for $K<1$.

\section{\label{PsiC}Calculating Phase Fluctuations from the Fixed-amplitude Models}
We start with the truncated model (\ref{Total1D1D}). For
convenience, we rewrite it as
\begin{eqnarray}
 S[\Upsilon]/\hbar&=&{1\over 2}\sum_{\vec{q}}\left[G_{\phi}^{-1}(\vec{q})\phi(\vec{q})\phi(-\vec{q})
 +G_{\theta}^{-1}(\vec{q})\theta(\vec{q})\theta(-\vec{q})\right]\nonumber\\
 &&-2g_2\psi_0\int dx d\tau\ \ \cos\left(2\phi-\theta\right)
 \label{}\, ,
\end{eqnarray}
where
\begin{eqnarray}
G_{\phi}^{-1}(\vec{q})&\equiv&{1\over\pi K}(q_x^2+q_{\tau}^2)\, ,\nonumber\\
G_{\theta}^{-1}(\vec{q})&\equiv&\left[ g_1^2\left(1\over
2n_e\right)^2q_x^2+q_{\tau}^2\right]\psi_0^2\, .
\end{eqnarray}
Then we define two new variables, so that the action in terms of
these two variables are decoupled. That is, the action can be
expressed as
\begin{eqnarray}
 S[\Upsilon, \Sigma]/\hbar&=&{1\over 2}\sum_{\vec{q}}G_{\Sigma}^{-1}(\vec{q})\Sigma(\vec{q})\Sigma(-\vec{q})
 +{1\over
 2}\sum_{\vec{q}}G_{\Upsilon}^{-1}(\vec{q})\times\nonumber\\
 &&\Upsilon(\vec{q})\Upsilon(-\vec{q})-2g_2\psi_0\int dx d\tau\ \ \cos{\Upsilon}\, ,\nonumber\\
 \label{Irrelevance}
\end{eqnarray}
where
\begin{eqnarray}
 \Upsilon(\vec{q})&\equiv&2\phi(\vec{q})-\theta(\vec{q})\, ,\label{Upsilon}\\
 \Sigma(\vec{q})&\equiv&\phi(\vec{q})-{2\Upsilon(\vec{q}) G_{\theta}^{-1}(\vec{q})\over G_{\phi}^{-1}(\vec{q})+4G_{\theta}^{-1}(\vec{q})}\, ,
 \label{Sigma}\\
 G_{\Upsilon}^{-1}(\vec{q})&\equiv&{G_{\phi}^{-1}(\vec{q})G_{\theta}^{-1}(\vec{q})\over G_{\phi}^{-1}
 (\vec{q})+4G_{\theta}^{-1}(\vec{q})} \, ,\label{PropagatorUpsilon}\\
 G_{\Sigma}^{-1}(\vec{q})&\equiv&G_{\phi}^{-1}(\vec{q})+4G_{\theta}^{-1}(\vec{q})\label{PropagatorSigma}\,
 .
\end{eqnarray}
Since $\Upsilon$ and $\Sigma$ are decoupled, using (\ref{Upsilon})
and (\ref{Sigma}) the fluctuations of $\theta$ and $\phi$ can be
calculated as
\begin{eqnarray}
 \langle\phi(\vec{q})\phi(-\vec{q})\rangle&=&G_{\Sigma}(\vec{q})
 +{4 G_{\theta}^{-2}(\vec{q})\langle\Upsilon(\vec{q})\Upsilon(-\vec{q})\rangle\over \left[G_{\phi}^{-1}
 (\vec{q})+4G_{\theta}^{-1}(\vec{q})\right]^2}\, ,\nonumber\\
 \label{PhiFluctuations}\\
 \langle\theta(\vec{q})\theta(-\vec{q})\rangle&=&4G_{\Sigma}(\vec{q})
 +{G_{\phi}^{-2}(\vec{q})\langle\Upsilon(\vec{q})\Upsilon(-\vec{q})\rangle\over \left[G_{\phi}^{-1}
 (\vec{q})+4G_{\theta}^{-1}(\vec{q})\right]^2}\, .\nonumber\\ \label{ThetaFluctuations}
\end{eqnarray}
To calculate $\langle\Upsilon(\vec{q})\Upsilon(-\vec{q})\rangle$, we
need to know whether the cosine term is relevant or irrelevant.
Thus, we evaluate
\begin{eqnarray}
 \int dx d\tau\ \ \langle\cos{\Upsilon}\rangle_0=\int dx d\tau\ \ e^{-\langle\Upsilon^2\rangle_0\over 2}\, ,
 \label{CosineAverage}
\end{eqnarray}
where $\langle\rangle_0$ is the average with respect to the harmonic
part of the action (\ref{Irrelevance}). Let us first calculate
%\begin{eqnarray}
% \langle\cos{\Upsilon}\rangle_0=e^{-{1\over
% 2}\langle\Upsilon^2\rangle_0}
%\end{eqnarray}
%where
\begin{eqnarray}
 \langle\Upsilon^2\rangle_0&=&{1\over (2\pi)^2}\int dq_xdq_{\tau}\ \
 G_{\Upsilon}(\vec{q})\nonumber\\
 &=&{1\over (2\pi)^2}\int dq_xdq_{\tau}\ \ \left\{{1\over\psi_0^2\left[q_{\tau}^2+g_1^2\left(1\over 2n_e\right)^2q_x^2\right]}
 +\right.\nonumber\\
 &&\left.{4\pi K\over {q_x^2+q_{\tau}^2}}\right\}\nonumber\\
 &=&\left({n_e\over \pi g_1\psi_0^2}+2K\right)\ln{L}\, .
 \end{eqnarray}
Plugging it into (\ref{CosineAverage}) we obtain
\begin{eqnarray}
\int dx d\tau\ \ \langle\cos{\Upsilon}\rangle_0\sim L^{2-K-{n_e\over
2\pi g_1\psi_0^2}}\, ,
\end{eqnarray}
%To simplify the integrant we divide the integral region into several
%parts. From now on we assume $\psi_0>\psi_0^c$ and the condition
%(\ref{ImplicitCondition}) always. For non-zero $K<1$ we can
%calculate the integral as the following
% \begin{eqnarray}
% &&\int dq_xdq_{\tau}\ \ G_{\Upsilon}(\vec{q})\nonumber\\
% &\simeq&{1\over 2\pi^2}\int_{-a/\alpha}^{a/\alpha} dq_x
%\int_0^{g_1q_x}dq_{\tau} {1\over g_1^2\psi_0^2\left(1\over
%2n_e\right)^2(q_x^2+q_\tau^2)}\nonumber\\
%&&+{1\over 2\pi^2}\int_{-a/\alpha}^{a/\alpha}
%dq_x\int_{g_1q_x}^{\psi_0^{-1}q_x}dq_\tau\ \
%{q_x^2\over \psi_0^2q_\tau^2(q_x^2+q_\tau^2)}\nonumber\\
%&&+{2K\over \pi}\int_{-a/\alpha}^{a/\alpha} dq_x\int_{\psi_0^{-1}q_x}^{\infty}dq_\tau\ \ {1\over(q_x^2+q_\tau^2)}\nonumber\\
%&\simeq&({C\over g_1\psi_0^2}+2K)\ln{L}
%\end{eqnarray}
%where $L$ is the length of the chains, $C$ some positive order 1
%constant.
which implies a critical line in the $\psi_0^{-2}$-$K$ plane defined
by
\begin{eqnarray}
 {n_e\over\pi g_1\psi_0^2}+2K=4\, .
 \label{ACriticalLine}
\end{eqnarray}
This critical line is illustrated in Fig. \ref{fig: CriticalLine}.
The cosine term in the action (\ref{Irrelevance}) is relevant below
the critical line and irrelevant above it. In the two limiting cases
the function (\ref{ACriticalLine}) predicts that $\psi_0^c\to\infty$
for $K=2$ and $\psi_0^c\sim g_1^{-1/2}$ for $K=0$. For $K<1$, it
predicts
\begin{eqnarray}
\psi_0^c\sim g_1^{-1/2}\, .
\end{eqnarray}
For $\psi_0<\psi_0^c$, the cosine term is irrelevant, and we get
$\langle\Upsilon(\vec{q})\Upsilon(-\vec{q})\rangle=G_{\Upsilon}(\vec{q})$,
which combined with (\ref{PhiFluctuations}) and
(\ref{ThetaFluctuations}) leads to
\begin{eqnarray}
 \langle\phi(\vec{q})\phi(-\vec{q})\rangle&=&G_{\phi}(\vec{q})\, ,\\
 \langle\theta(\vec{q})\theta(-\vec{q})\rangle&=&G_{\theta}(\vec{q})\,
 .
\end{eqnarray}
This implies that $\theta$ and $\phi$ are effectively decoupled.
For $\psi_0>\psi_0^c$,
$\langle\Upsilon(\vec{q})\Upsilon(-\vec{q})\rangle$ is massive, and
we obtain
\begin{eqnarray}
 \langle\phi(\vec{q})\phi(-\vec{q})\rangle&=&G_{\Sigma}(\vec{q})\, ,\label{BindPhi}\\
 \langle\theta(\vec{q})\theta(-\vec{q})\rangle&=&4G_{\Sigma}(\vec{q})\,
 ,\label{BindTheta}
\end{eqnarray}
which implies that the fluctuations of $\theta$ and $\phi$ are bound
together.

Now we discuss the model (\ref{Total3D3}). We follow the same
routine as the above. Defining $\Upsilon$ and $\Sigma$ using
(\ref{Upsilon}) and (\ref{Sigma}), we get a decoupled action:
\begin{eqnarray}
 S[\Upsilon, \Sigma]/\hbar&=&{1\over 2}\sum_{\vec{q}}G_{\Sigma}^{-1}(\vec{q})\Sigma(\vec{q})\Sigma(-\vec{q})+
 {1\over 2}\sum_{\vec{q}}G_{\Upsilon}^{-1}(\vec{q})\times\nonumber\\
 &&\Upsilon(\vec{q})\Upsilon(-\vec{q})
 -{2g_2\psi_0\over\pi^2}\int dx d^2r_{\perp} d\tau\ \
 \cos{\Upsilon}\, ,\nonumber\\
 \label{}
\end{eqnarray}
where $G_{\Upsilon}^{-1}(\vec{q})$ and $G_{\Sigma}^{-1}(\vec{q})$
are defined in (\ref{PropagatorUpsilon}) and
(\ref{PropagatorSigma}), respectively. $G_{\phi}^{-1}(\vec{q})$ and
$G_{\theta}^{-1}(\vec{q})$ are now defined as
\begin{eqnarray}
G_{\phi}^{-1}(\vec{q})&\equiv&{1\over\pi^3 K}(q_x^2+q_{\tau}^2)\, ,\\
G_{\theta}^{-1}(\vec{q})&\equiv&\left(\psi_0\over \pi\right)^2\left[
g_1^2\left(1\over
2n_e\right)^2q_x^2+q_{\tau}^2+C_{\perp}q_{\perp}^2\right]\,
.\nonumber\\
\end{eqnarray}
To check whether the cosine term is relevant or not,
%and then obtaining an
%action only involves $\Upsilon$:
%\begin{eqnarray}
% S[\Upsilon]/\hbar&=&{1\over 2}\sum_{\vec{q}}G_{\Upsilon}^{-1}(\vec{q})\Upsilon(\vec{q})\Upsilon(-\vec{q})
% -{2g_2\psi_0\over\pi^2}\times\nonumber\\
% &&\int dxd^2r_{\perp} d\tau\ \ \cos{\Upsilon}\, ,\nonumber\\
% \label{Irrelevance3D}
%\end{eqnarray}
%where $G_{\Upsilon}^{-1}(\vec{q})$ is different from
%(\ref{PropagatorUpsilon1D}), given by
%\begin{eqnarray}
% G_{\Upsilon}^{-1}(\vec{q})={2\psi_0^2(q_x^2+q_{\tau}^2)\left[q_{\tau}^2+g_1^2\left(1\over 2n_e\right)^2
% q_x^2+C_{\perp}q_{\perp}^2\right]\over{(8\pi K\psi_0^2+2)q_{\tau}^2+\left[8\pi
% K\psi_0^2g_1^2\left(1\over 2n_e\right)^2+2\right]q_x^2}+8\pi K\psi_0^2C_{\perp}q_{\perp}^2}\, .\nonumber\\
%\label{PropagatorUpsilon3D}
%\end{eqnarray}
%Then
we evaluate the average of the cosine term for a single chain with
respect to the harmonic part of the action. After we
calculate
\begin{eqnarray}
 \langle\Upsilon^2\rangle_0&=&{1\over (2\pi)^4}\int dq_xd^2q_{\perp}dq_{\tau}\ \
 G_{\Upsilon}(\vec{q})\nonumber\\
 &=&{1\over (2\pi)^4}\int dq_xd^2q_{\perp}dq_{\tau}\ \ \left\{
 {4\pi^3 K\over {q_x^2+q_{\tau}^2}}+\right.\nonumber\\
 &&\left.{\pi^2\over\psi_0^2\left[q_{\tau}^2+g_1^2\left(1\over 2n_e\right)^2q_x^2+C_{\perp}q_{\perp}^2\right]}\right\}\nonumber\\
 &\sim& 2K\ln{L}\, ,
\end{eqnarray}
we obtain
\begin{eqnarray}
\int dx d\tau\ \ \langle\cos{\Upsilon}\rangle_0\sim L^{2-K}\, .
\end{eqnarray}
This implies that for $K<2$, the cosine term in the action
(\ref{Total3D3}) is always relevant no matter the value of $\psi_0$.
Therefore, for $K<1$ and arbitrary $\psi_0$, the action
(\ref{Total3D3}) predicts that the fluctuations of $\phi$ and
$\theta$ are bound together and given by (\ref{BindPhi}) and
(\ref{BindTheta}), respectively.


\begin{thebibliography}{99}
\bibitem{Smart ass}
Seen from a point with a distance larger than $1/2K_F$ from the
central chain, the $2K_F$ modulation is even out. So the system is
essencially a cylinder with radius $1/2K_F$ centered at the central
chain. Therefore, it is effectively one-dimensional.
\end{thebibliography}
\end{document}